\documentclass[12pt,preprint]{aastex}

\def\cm#1{\ifmmode {\,{\rm cm^{-#1}}}                  % cm-1, cm-2, cm-3, ...
        \else \hbox{$\,${\rm cm$^{\rm -#1}$}}\fi}

\newcommand{\irase}{IRAS\,18276$-$1431} %% irase-ighteen
\newcommand{\doce}{\mbox{$^{12}$CO}}
\newcommand{\trece}{\mbox{$^{13}$CO}}
\newcommand{\kms}{\mbox{km~s$^{-1}$}}

\newcommand{\mloss}{\mbox{$\dot{M}$}}
\newcommand{\my}{\mbox{$M_{\odot}$~yr$^{-1}$}}
\newcommand{\ls}{\mbox{$L_{\odot}$}}
\newcommand{\msun}{\mbox{$M_{\odot}$}}

\newcommand{\ai}{\mbox{\'{\i}}} 
\newcommand{\vexp}{\mbox{$V_{\rm exp}$}} 
\newcommand{\vsys}{\mbox{$V_{\rm sys}$}} 
 
\newcommand{\h}{$^{\rm h}$}
\newcommand{\m}{$^{\rm m}$}

\newcommand{\tex}{\mbox{$T_{\rm ex}$}}
\newcommand{\teff}{\mbox{$T_{\rm eff}$}}
\newcommand{\tkin}{\mbox{$T_{\rm kin}$}}

\newcommand{\kp}{\mbox{$K_{\rm p}$}}
\newcommand{\lp}{\mbox{$L_{\rm p}$}}
\newcommand{\ms}{\mbox{$M_{\rm s}$}}
\newcommand{\hstcol}{\mbox{F814W$-$F606W}}

\shorttitle{Adaptive Optics Imaging of IRAS\,18276-1431}
\shortauthors{S\'anchez Contreras et al.}

\begin{document}

\title{Adaptive Optics Imaging of \irase: 
a bipolar pre-planetary nebula with circumstellar ``searchlight beams'' and ``arcs''}

\author{C. S\'anchez Contreras\altaffilmark{1}, D. Le Mignant\altaffilmark{2}, 
R. Sahai\altaffilmark{3}, A. Gil de Paz\altaffilmark{4}, and M. Morris\altaffilmark{5}}

%\altaffiltext{1}{California Institute of Technology, MS\,105-24, 
%Pasadena, CA 91125}

\altaffiltext{1}{Departamento de Astrof\ai sica Molecular e Infrarroja, Instituto de Estructura de la Materia, CSIC, Serrano 121, E-28006 Madrid, Spain}

\altaffiltext{2}{W. M. Keck Observatory, 65-1120 Mamalahoa Highway,
Kamuela, HI 96743, USA}

\altaffiltext{3}{Jet Propulsion Laboratory, MS 183-900, California
Institute of Technology, Pasadena, CA 91109, USA}

\altaffiltext{4}{Departamento de Astrof\ai sica, Universidad Complutense de Madrid, Avda. de la Complutense S/N, E-28040, Madrid, Spain}

\altaffiltext{5}{Department of Physics \& Astronomy, University of California Los Angeles, Box 951547, Los Angeles, CA 90095-1547, USA}

%% \email{carmen@damir.iem.csic.es}


\begin{abstract}
We present high-angular resolution images of the post-AGB nebula
\irase\ (also known as OH 17.7-2.0) obtained with the Keck\,II
Adaptive Optics (AO) system in its Natural Guide Star (NGS) mode in
the \kp, \lp, and \ms\ near-infrared bands. 
%High angular resolution
%is needed to spatially resolve the nebular structure of post-AGB
%objects and to make progress in our understanding of their evolution.
We also present supporting optical F606W and F814W $HST$ images as
well as interferometric observations of the \doce\,($J$=1--0),
\trece\,($J$=1--0), and 2.6\,mm continuum emission with OVRO.  The
envelope of \irase\ displays a clear bipolar morphology in our optical
and NIR images with two lobes separated by a dark waist and surrounded
by a faint 4\farcs5$\times$3\farcs4 halo.  Our \kp-band image reveals
two pairs of radial ``searchlight beams'' emerging from the nebula
center and several intersecting, arc-like features.  From our CO data
we derive a mass of $M>$0.38\,[$D$/3\,kpc]$^2$\,\msun\ and an
expansion velocity
\vexp=17\,\kms\ for the molecular envelope.  The density in the halo
follows a radial power-law $\propto$$r^{-3}$, which is consistent with
a mass-loss rate increasing with time.  Analysis of the NIR colors
indicates the presence of a compact central source of $\sim$300-500\,K
dust illuminating the nebula in addition to the central star.
Modeling of the thermal IR suggests a two-shell structure in the dust
envelope: 1) an outer shell with inner and outer radius $R_{\rm
in}$$\sim$1.6$\times$$10^{16}$cm and $R_{\rm
out}$$\ga$1.25$\times$10$^{17}$cm, dust temperature $T_{\rm
d}$$\sim$105-50\,K, and a mean mass-loss rate of
\mloss$\approx$10$^{-3}$\my; and 2) an inner shell with $R_{\rm
in}$$\sim$6.3$\times$10$^{14}$\,cm, $T_{\rm dust}$$\sim$500-105\,K, and
\mloss$\sim$3$\times$10$^{-5}$\my. 
An additional population of big dust grains (radius $a$$\ga$0.4\,mm) with
$T_{\rm dust}$=150-20\,K and mass $M_{\rm
dust}$=(0.16-1.6)$\times$10$^{-3}$[$D$/3\,kpc]$^2$\msun\, can account for
the observed sub-mm and mm flux excess. The mass of the envelope
enclosed within $R_{\rm out}$=1.25$\times$10$^{17}$cm derived from SED
modeling is $\sim$1[$D$/3\,kpc]$^2$\msun.
\end{abstract}





\section{Introduction}
\label{intro}

In recent years, $\sim$0\farcs1-resolution imaging with $HST$ has
provided many new insights into the evolution from the Asymptotic
Giant Branch (AGB) to the phase of Planetary Nebula (PN).  This
transition, which takes place through the intermediate short-lived
($\sim$10$^3$\,yr) stage of pre-planetary nebula (PPN), is
characterized by dramatic changes of the nebular morphology and
kinematics: the spherical, slowly expanding ($V_{\rm
exp}$\,$\sim$\,15\,\kms) AGB circumstellar envelope (CSE) develops
clear departures from sphericity and fast ($\ga$\,100\,\kms) outflows
directed in opposing directions along one or more axes.  Although
there is no consensus yet for what causes this spectacular
metamorphosis, fast jet-like winds have been hypothesized to play an
important role (see e.g.\ a recent review by Balick \& Frank,
2002). In light of the many structural details revealed by $HST$ data,
e.g., the multiple lobes and the high degree of point symmetry of
outflows observed in most young PNs, and PPNs, Sahai \& Trauger (1998)
hypothesized that fast collimated outflows are the primary agent for
setting the stage in the shaping of PNs, rather than a pre-existing
equatorial density enhancement in the AGB CSEs, as assumed in the
popular Generalized Interacting Stellar Winds model
\citep{bal87}.  These outflows carve out a complex imprint within
the AGB CSE producing and shaping the fast, bipolar lobes observed in
most PPNs and PNs.

In order to study the poorly known mass-loss processes mediating the
AGB-to-PN transition, we have obtained 1.6-4.7\,\micron\ images of
PPNs with the Keck Adaptive Optics (AO) system (S\'anchez Contreras et
al., 2006). This forms part of our extensive, multi-wavelength studies
of PPNs, which also include optical/NIR $HST$ imaging and optical
spectroscopy \cite{sah04} and millimeter wavelength interferometry
\cite{san04}.  Progress in our understanding of post-AGB evolution
requires high spatial resolution because most PPNs are very small in
angular extent ($\la$5\arcsec), exhibit rich structure at a scale
$\la$0\farcs1, and because the shaping mechanism operates very close
to the central star, $\la$10$^{16}$cm (Balick \& Frank 2002, S\'anchez
Contreras \& Sahai 2001). Also, observing at long wavelengths is
desirable, since the dusty CSEs of most post-AGB objects are very
optically thick.

In this paper we present \kp, \lp, and \ms\ AO images of the PPN
\irase, together with optical WFPC2/$HST$ and interferometric 2.6\,mm
(CO line and continuum emission) data, which have helped our
understanding of the complex structure of this object. Observational
details and results are described in Sections \ref{obs} and
\ref{results}. The analysis of the NIR and optical 
colors to derive the dust extinction towards the central star and the
nature of the illuminating source are presented in Sections
\ref{extinction} and \ref{illum}. Structural components and physical 
parameters (mass, mass-loss rate, and dust temperature) of \irase\ are
discussed in Section \ref{analysis} and \ref{phys}. The formation and
evolution of \irase\ is described in Section \ref{discuss}. Our main
conclusions are summarized in Section \ref{summary}.

\subsection{The pre-planetary nebula \irase}
\label{obj}

\irase\ (also known as OH 17.7-2.0) is an OH/IR star 
that has reached the end of its evolution on the AGB 
and is rapidly evolving to the PN phase.
%There are many observational indications of the post-AGB nature of
%this object. 
\irase\  displays the classical bimodal spectral energy distribution (SED) 
observed in PPNs, which is indicative of a detached CSE
with relatively cool ($\sim$150\,K) dust grains
%% with $\sim$150\,K dust 
\cite{LeBertre87}. The progressive disappearance of H$_2$O maser emission 
in \irase\ is consistent with a recent drop of the mass-loss rate from
\mloss$\ga$10$^{-5}$ to well below \mloss=10$^{-7}$\,\my\ \cite{eng02}. 
The cessation of the large scale mass-loss, indicated by both the
disappearance of the H$_2$O masers and the presence of a detached
envelope, is believed to designate the beginning of the post-AGB
evolution.  The non-detection of SiO masers in \irase\ \cite{Nym98} is
also consistent with the post-AGB nature of this object.  The central
star of \irase\ has a spectral type earlier than $\sim$K5, with an
effective temperature in the range
\teff=4000-10,000\,K, and a total luminosity of
$L\sim$1.5$\times$10$^4$\,[$D$/4.6kpc]$^2$ (Le Bertre et al$.$ 1989). 
This is a confirmation that the object has left the AGB.
%% the luminosity could be underestimated from the SED. This is because 
%% for an almost edge-on nebula, like iras18276, only a very small fraction 
%% of the optical photons escaping along the polar direction are scattered 
%% towards us.

Ground-based imaging polarimetry in the $J$ and $K$ near-infrared
bands shows an extended, non-spherical scattering envelope around
\irase\ \cite{gled05}. 
The angular size of the envelope and position angle of its major axis
in the $J$ ($K$) band are $\sim$3\arcsec$\times$2\arcsec\ and
PA=24$\pm$2\degr\ (3\farcs9$\times$3\farcs4 and PA=22$\pm$2\degr),
respectively. The polarized intensity images reveal a clear bipolar
morphology and polarization vectors that are mostly aligned
perpendicular to the bipolar axis, at PA$\approx$23\degr. This is
typical of scattering in an optically thick envelope, where light
escapes preferentially along bipolar cavities
\cite{bai03,gled05}. The axisymmetric scattering models by Gledhill (2005) suggest that the lobes of \irase\ lie very close to the plane of the sky, with an approximate inclination of $i$$\sim$3\degr.
%(Photometry, J-K=3.4mag). 
%The envelope of \irase\ is marginally resolved in the 9.8\micron\ 
%image reported by Meixner et al. (1999). 

Strong OH maser emission is observed in the equatorial regions
\irase\, (Bains al.\ 2003).  This dense, equatorial component is
elongated in the direction of PA$\sim$110\degr, i.e., is approximately
perpendicular to the polar cavities, and has inner and outer angular
radii of 0\farcs25 and 0\farcs85, respectively. As discussed by Bains
et al. (2003), horn-like features observed in the OH maser
position-velocity diagram (see their Fig.\,10) probably trace the base
of a bipolar outflow very close to the central source. The relative
absence of OH masers in the polar regions (compared with the much more
intense OH emission observed in the equatorial regions) is consistent
with the interpretation of these polar regions as cavities within
which the lower gas density fails to generate strong maser
emission. These OH maser observations also suggest a large-scale,
regular magnetic field of strength B=+4.6mG (1612MHz) and B=+2.5mG
(1667MHz) in \irase\ that could have played a role in causing the
equatorially enhanced mass loss in this object. A burst of strongly
polarized OH maser emission that could be linked to recent
evolutionary changes in the shell has been reported by Szymczak
\& G{\'e}rard (2005).
%%\citep{szy05}. 
%The molecular envelope of \irase\ has also been studied from a single-dish 
%\doce$J$=2--1 spectrum, which indicates an expansion velocity of 12.2\kms\ and 
%a mass-loss rate of \mloss=0.58$\times$10$^{-5}$\,\msun\
%\cite[][for a distance of 2kpc]{heske90}.

The distance to \irase\ is uncertain. OH phase-lag measurements
indicate a distance in the range $D$=2-5.4\,kpc \cite{bow83,her85}.
The near kinematic distance is $D$=4.6\,kpc \cite{LeBertre89}. In this
paper, we adopt an intermediate value of $D$=3\,kpc. 
%%, which implies a total luminosity of $L$=6400\,\ls.

\section{Observations}
\label{obs}

\subsection{Near-Infrared Natural Guide Star AO imaging}

We have imaged the dust envelope around the OH/IR star \irase\ through
the K$_p$(2.12\micron), L$_p$(3.8\micron) and M$_s$(4.7\micron)
filters with the W. M. Keck II 10-m telescope using the NIRC2 Camera
(Matthews et al. 2005, in prep) with Adaptive Optics (AO) (Wizinowich
et al. 2000). \irase\ was observed with AO in its Natural Guide Star
(NGS) mode on UT 2005 August 12 at airmass 1.22--1.26.
%% UT 6:30-6:56 airmass 1.22-1.26
The AO guide star, USN-0755-0509985 (m$_V$=11.0) is located 23\farcs2 S and
0\farcs13 E from \irase. 
%  Vmag = 11.0 WFS speed at 315Hz, 110 counts. 
The AO system ran at the speed of 315\,Hz
with an average intensity per sub-aperture of 120\,counts. 
The field of view (FoV) and pixel scale of the
NIRC2 narrow camera is $10''\times10''$ and 9.94 mas pixel$^{-1}$.
The seeing was 0\farcs4 in the visible as
measured on the acquisition camera.

The images in the different bands were acquired 
using a generic 3-position dithering script avoiding the noisiest 
quadrant of NIRC2 where the number 
of frames for each position was 3 in \kp, 6 in \lp\, and 6 in \ms. 
The images have been reduced in a standard way. First,
we derived a sky frame by computing the filtered 
median of all frames for each band. Then, each frame was
sky-subtracted, flat-fielded, corrected for bad pixels and
shift-and-added in a final image for each observing setup. 
We also observed the photometric standard \#148 S810-A from
%% {\bf (FS\#148 - 19:41 23.3 ...)}
Leggett et al. (2003) at UT 08:30 in the same region of the sky 
that we used to derive the photometry zero points 
for \kp, \lp, and \ms.
%% kp 1% 
%% Lp 3%
%% Ms 10% 
%{\bf L$_p$
%and M$_s$ images have been flux calibrated using the photometric
%standard XXXXXXX (Leggett et al. 2003) on UT Aug 12. 
We have used the magnitude-to-mJy flux density conversion factors
given by Tokunaga and Vacca (2005).

%Standards are from UKIRT faint standard. I used A0 to
%double-check fluxes between standards.
%148  s810-a       19 41 23.52 -03 50 56.1    
%149  p338-c       20 00 39.25 +29 58 40.0    
%Legett et al. 2003, MNRAS, 345, 144
%Extinction coefficients, alpha are from UKIRT as well.
%  see Krisciunas et al., PASP 99, 887, 1987
%FS#148      A0 star
%------
%S810 a 19 41 23.5 -03 50 56.1 2000.0  
%Observed at UT 08:30 

The spatial resolution in the images is nearly diffraction limited for
the Keck II telescope in \kp\ (60-65 versus 48 mas) and
diffraction-limited in \lp\ and \ms\ (Table 1).  The PSF of our images
has the typical elongation resulting from the AO anisoplanatism
effect, as expected when the AO guide star is offset from the science
target.  This effect is seen in the \kp\ band images where the PSF
varies from 55\,mas near the AO guide star to 75\,mas in the most
distant imaged stars. This elongation is not seen in \lp\ and
\ms\ as the isoplanatic angle becomes much larger ($\ge$\,50\arcsec).
We have used the nearest stars to \irase\ to check the image quality
and characterize the PSF in our target (Table 1).  The Strehl ratio
(SR) was measured to be 0.16 (0.75) in the \kp\ (\lp) image, giving a
scaled SR of 0.80 in \ms.

%% PSF - halo?? 

\subsection{Optical F606W and F814W $HST$ imaging}

\irase\ was imaged on UT date 2002 Feb 16 (GO/SNAP program 9101) with the
800$\times$800-pixel planetary camera of the Wide-Field \& Planetary Camera
2 (WFPC2), which has a plate scale of 0\farcs0456/pixel, using the F606W
[$\lambda=0.60\micron$, $\Delta\lambda=0.123\micron$; exposures were
2$\times$5\,s and 2$\times$100\,s] \& F814W [$\lambda=0.80\micron$,
$\Delta\lambda=0.149\micron$; exposures were 2$\times$3 \& 2$\times$60\,s]
filters. For both filters, each of the two long-exposure images were
dithered relative to each other by about 7.7 pixels to improve the image
quality. The standard pipeline processed images were retrieved from the $HST$
archive. The short-exposure images were used to check for saturated pixels
in the long-exposure ones, and none were found. The long-exposure images were
subsampled by a factor 2, and then registered to a common reference frame
with high precision (i.e., to within $\lesssim$0.1 pixel) using field
stars. Cosmic-ray removal and averaging was then carried out using the
registered images.

\subsection{Millimeter-wavelength mapping with OVRO}

Interferometric mapping of the \mbox{$^{12}$CO\,$J$=1--0} and
\mbox{$^{13}$CO\,$J$=1--0} emission lines at 115 and 110\,GHz of
\irase\ was carried out using the millimeter interferometer of the
Owens Valley Radio Observatory (OVRO\footnote{\tt
http://www.ovro.caltech.edu/mm}) on Sept 29, 2002,
with 5 antennas and baselines ranging from 18\,m to 55\,m.  The units
of the digital spectral line correlator were arranged to provide a
total bandwidth of 90\,MHz ($\sim$\,234\,\kms) with a channel spacing
of 1\,MHz (corresponding to $\sim$2.6\,\kms). 
%% C18O was also observed but 
%% not detected - the line peak is expected at Vlsr=60km/s, 
%% that is at the edge of the spectral window chosen: centered at Vlsr=0kms with bandwith=78km/s.
The 3\,mm continuum emission from
\irase\ was observed simultaneously using the dual-channel analog
continuum correlator, which provided a total bandwidth of 4\,GHz (after
combining both IF bands).  Total duration of the track (including
integration time on source and calibrators) was $\sim$\,6hr.
\irase\ was observed $\pm$2.6\,hr around its transit.

The calibration of the data was performed using the MMA software
package\footnote{MMA is written and maintained by the Caltech's
Millimeter Interferometry Group.}.  Data were gain calibrated in
baseline-based mode using the quasar J1833$-$210, which was observed at
regular time intervals of $\sim$\,20\,minutes before and after our
target. The quasar 3C\,273 and the planet Uranus were used as passband
and flux calibrators, respectively. The flux of the gain calibrator
J1833$-$210 was estimated to be 2.2\,Jy at 115\,GHz. Flux calibration
errors could range up to 30\%.

Reconstruction of the maps from the visibilities was done using the
Multichannel Image Reconstruction, Image Analysis and Display (MIRIAD)
software. We Fourier transformed the measured visibilities with robust
weighting, which is an optimized compromise between natural and
uniform weighting, for \doce\ and natural weighting for \trece\ for
S/N optimization. After that, data were cleaned and maps restored.
The clean beam for our \doce\ (\trece) maps has
FWHM=10\farcs9$\times$7\farcs1 (FWHM=11\farcs5$\times$8\farcs7) and is
oriented at PA=73\degr\ (PA=58\degr). The noise (1$\sigma$) in our
\doce\ (\trece) maps, as measured in channels with no signal,
is 0.12 (0.08)\,Jy\,beam$^{-1}$. 
%In the central channels, where the \doce\ (\trece)
%intensity peaks, the noise is 0.13 (xx)\,mJy\,beam$^{-1}$.
In our 3mm continuum maps $\sigma$=4\,mJy\,beam$^{-1}$.  The
conversion factor from \doce\ (\trece) surface brightness to
temperature units is 1.19 (1.01)\,K per Jy\,beam$^{-1}$.
The coordinates of the tracking center in our maps are
RA=18\h30\m30\fs75, Dec=$-$14\degr28\arcmin56\farcs28 (J2000.0).

\section{Observational results}
\label{results}

\subsection{Optical and NIR Imaging} %% HST and AO
\label{resimaging}

\begin{figure}
\epsscale{0.36}
\rotatebox{270}{\plotone{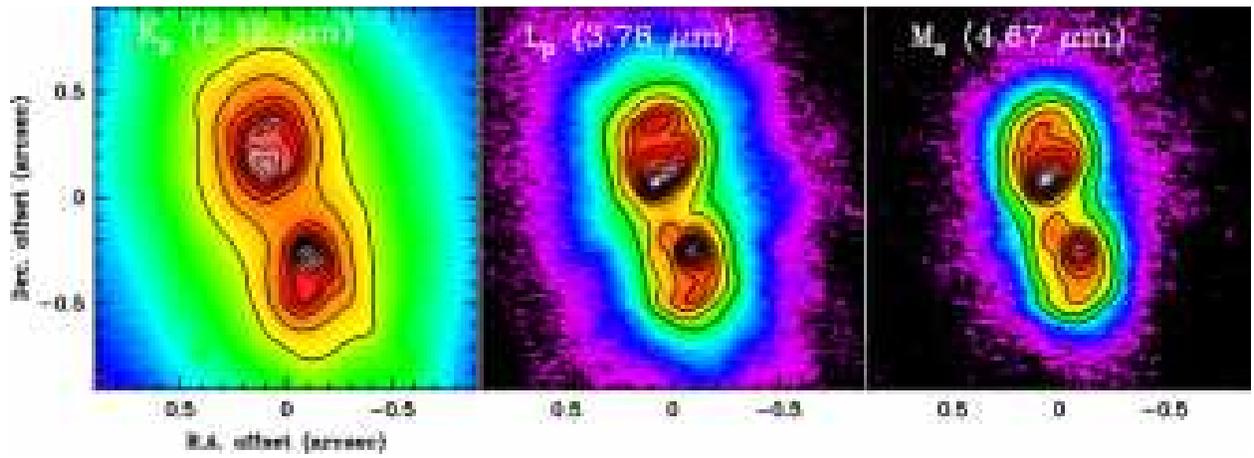}}
\caption{Adaptive optics images of \irase\ in the K$_{\rm p}$, 
L$_{\rm p}$, and M$_{\rm s}$ near-infrared bands. A logarithmic
intensity scale, from 3$\sigma$ to 95\% of the intensity maximum in
each image has been used (see Table\,\ref{tabphot}).  Contours are for
levels at 5\%, and from 10\% to 100\% of the maximum in each image in
10\% steps.}
\label{f1}
\end{figure}

\begin{figure}
\epsscale{0.5}
\rotatebox{270}{\plotone{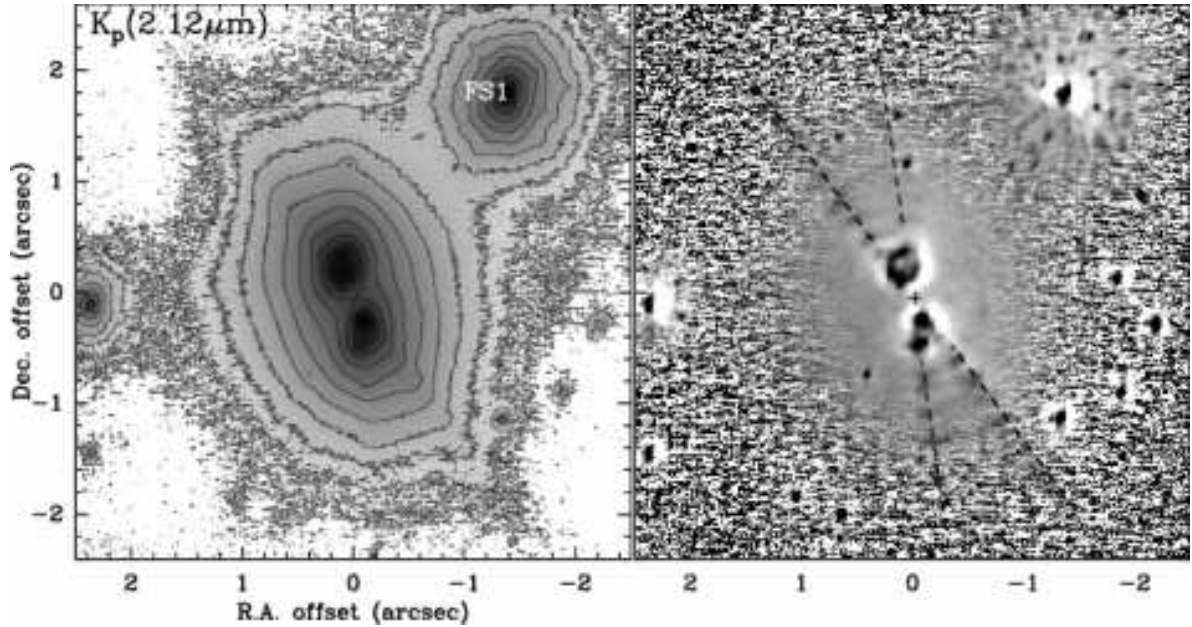}}
\caption{ {\it Left)} K$_{\rm p}$-band image of \irase\ showing the faint, elongated halo beyond the bipolar lobes. 
Intensity scale (shown in reverse gray scale) is the same as in
Fig.\ref{f1}. Contour levels are 3$\sigma$$\times$2$^i$ for $i$=0 to
11 ($\sigma$=8$\times$10$^{-6}$\,mJy\,pix$^{-1}$,
Table\,\ref{tabphot}).  The two bright objects at offsets
$-$1\farcs4,+1\farcs8 (referred to as FS1 in the text) and +2\farcs4,0
are not physically associated with \irase, as indicated by
polarization maps of Gledhill (2005). {\it Right)} A sharpened version
of the image on the left panel to show the ``searchlight beams'' and
``arcs'' of \irase. The maximum (black) and minimun (gray) values of
the displayed logarithmic intensity scale are set at 0.8 and 1.3,
respectively. The cross indicates the point where the searchlight
beams intersect each other. The center of the image (offset 0,0) is the center of the halo (see text in \S\,\ref{resimaging}). }
\label{f2}
\end{figure}

\begin{figure}
\epsscale{0.5}
%\plotone{I18276KpLpMs.ps}
%\plotone{false_color.ps}
\label{f4}
\begin{center}
%% \resizebox{0.7\hsize}{!}{\includegraphics*[108,208][503,603]{iras18276_fig3.ps}} 
\plotone{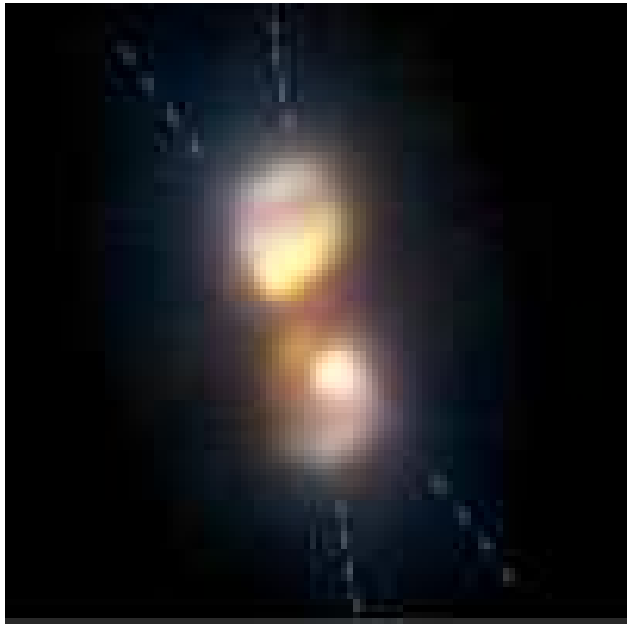}
\caption{A false three-color composite of the \kp\,(blue), \lp\,(green), 
and \ms\,(red) AO images in a squared-root intensity scale. Dashed
lines indicate the orientation of the searchlight beams. Note the well
delineated lobe caps and the equatorial waist of dust obscuring the
central star.  }
\end{center}
\end{figure}

\begin{figure}
\epsscale{0.34}
\begin{center}
\rotatebox{270}{\plotone{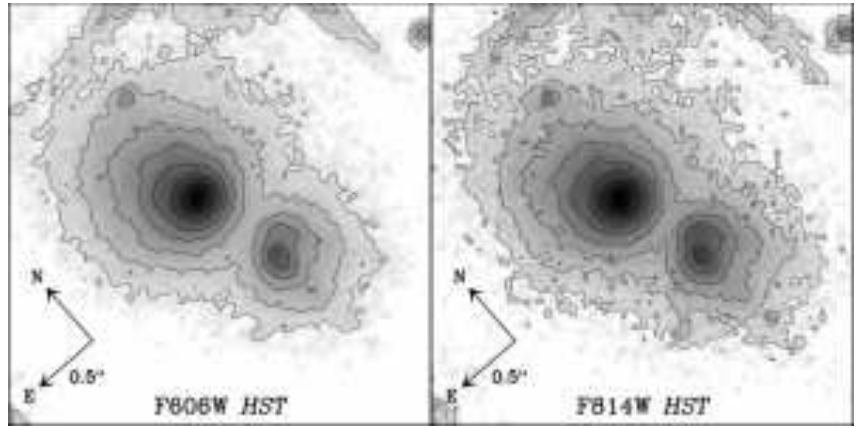}}
\caption{F606W and F814W WFPC2/$HST$ images of \irase\ in a logarithmic intensity scale.
Contour levels are 6$\sigma$$\times$2$^i$ for $i$=0 to 8
(%$\sigma$=4$\times$10$^{-6}$\,mJy\,pix$^{-1}$ for F606W and
%$\sigma$=10$^{-5}$\,mJy\,pix$^{-1}$ for F814W, 
see Table\,\ref{tabphot} for values of $\sigma$). 
The North and East directions are indicated by the 0\farcs5-sized arrows.}
\label{f3}
\end{center}
\end{figure}

The NIR and optical images of \irase\ are presented in
Figs$.$\,\ref{f1}-\ref{f3}.  
%A three color composite, made
%from the \kp, \lp, and \ms\ AO images is shown in Fig$.$\ref{f4}.
\irase\ displays a clear bipolar morphology, with two lobes oriented
along PA$\approx$17\degr-20\degr\ separated by a dark waist.  Our AO
images have higher angular resolution, probe deeper through the dusty
nebula, have larger dynamical range, and provide more structural
details than the $HST$ images. In particular, they show a sharp drop
of intensity at the lobe edges, which are quite well defined. The
lobes also have limb-brightened tips (or $caps$), suggesting that they
are closed-end structures. These caps are not visible in the optical
images. The overall size and morphology of the two lobes are
different, the SW lobe being shorter and narrower than the NE lobe.
%% length ~0.396\arcsec vs 0.36\arcsecs
%% give fwhm angular size.
The morphological differences between the lobes are most prominent at the
shortest wavelengths (optical and \kp-band images), where the SW lobe appears as a ``two-finger''
or $\wedge$-like structure in contrast with a more rounded NE lobe (Fig$.$\,\ref{f2} Right).
%% >>>>>>>>>>>>>>>>>>>>> brightness-related comments.... 
%{\it ??? The brightest region in the SW lobe (``P'') is observed at
%the same position in all the IR images.  Feature P gets progressively
%brighter with increasing wavelength in the IR.  The position of the
%brightness peak in the N lobe is not the same in all images.
%%%{\bf three peaks in the SW lobe?}
%The NE-to-SW lobe IR intensity ratio of the maximum peaks of the
%two lobes increases with wavelength being 0.7 in K$_{\rm p}$, 1.04 in L$_{\rm
%p}$, and 1.4 in M$_{\rm s}$. - BRIGHTNESS-RELATED COMMENTS ONLY IF WE CONCLUDE SOMETHING IMPORTANT LATER} 
% SO WHAT?? we can leave that sentence but NOT in the table. 

\begin{figure}
\epsscale{0.4}
\rotatebox{270}{\plotone{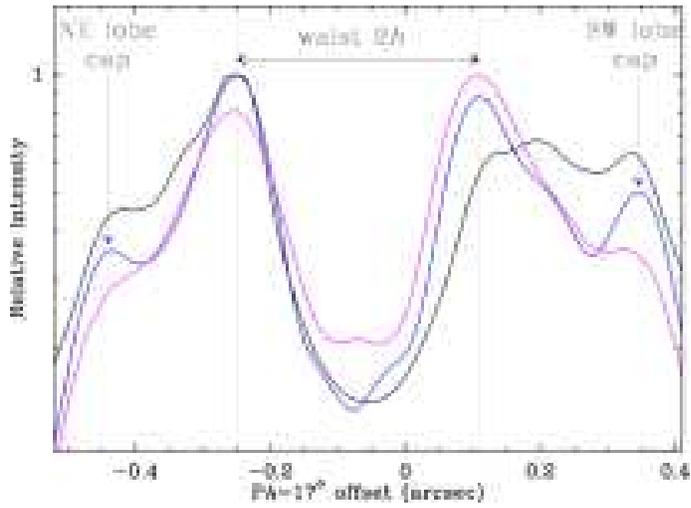}}
\caption{Radial profile through the lobes along PA=17\degr\ of the 
K$_{\rm p}$ (black), \lp\ (blue), and \ms\ (pink) images showing the
dark waist and the lobe caps.}
\label{cutlobe}
\end{figure}

%% >>>>>>  Waist 
The thickness of the dark equatorial waist between the lobes is larger
in the optical images than in the NIR ones.  This is because the
optical depth of the material in the nebular equator obscuring the
base of the lobes is expected to decrease with wavelength.  Radial
cuts of the NIR intensity along the lobes (Fig.\,\ref{cutlobe}) show a
central minimum, where the extinction reaches a maximum value, and two
adjacent absolute maxima (or peaks) separated by a distance
2$h$=0\farcs34 in the \lp\ and \ms\ images. (In the \kp-band image the
separation between the peaks is slightly larger presumably due to a
larger optical depth in \kp). The NIR optical depth along the line of
sight at the peaks (i.e. at $\pm h$ from the central minimum) is
expected to be $<<$1 given the sudden change in the slope of the
radial profile (from a positive to a negative value): we start to see
the base of the lobes through the waist. This probably indicates that
the optical depth and, therefore, the column density of the waist
decreases with latitude or, equivalently, with the distance above
the equatorial plane. A density power law that decreases with the
latitude is also inferred from modelling of polarimetric NIR data of
\irase\ by Gledhill (2005).

%% 38 pixels in Kp: 
%% ###### do the radial cuts also for the HST images and get the numbers. 
%% WAIST is 0.6!! 
%% ###### Parabolic shape?? or what?? need to see what the cut across and 
%% along the torus is telling us about the latitude-dependence of the density.

%% >>>>> Halo
Our deep NIR images also reveal a faint (sensitivity limited),
elongated halo surrounding the lobes. The contribution of the extended
PSF halo to this component is negligible, less than $\sim$1\%-2\%. The size of the halo
is 4\farcs5$\times$3\farcs4, as measured to a $\sim$3$\sigma$ level
over the sky background in our K$_{\rm p}$ image, and decreases with
increasing wavelength, suggesting that the halo consists of scattered
starlight.  The decrease of size with wavelength may be also partially
due to the lower sensitivity in the \lp- and \ms-band images. The
major axis of the halo is oriented along PA$\sim$17\degr, similarly to
the lobes as seen in the optical and NIR images. (In contrast, the extended PSF
halo, which is observed around bright field stars, is elongated
towards the guide star.) In the L$_{\rm p}$ image, there is faint
emission extended in the direction roughly perpendicular to the lobes,
along PA=105-110\degr. The outer radius of this sensitivity-limited
structure is 0\farcs6.
%% as measured in the \lp\ image 

\begin{figure}
\epsscale{0.4}
\plotone{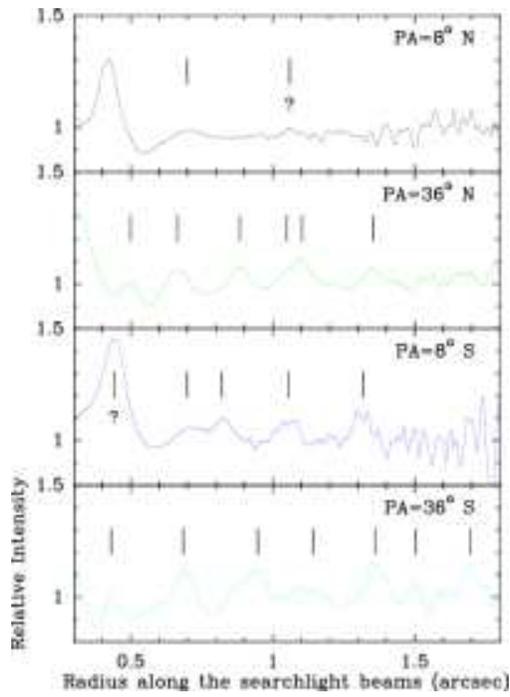}
\caption{Radial profile of the \kp-sharpened image (Fig.\,\ref{f2} Right) 
through the searchlight beams integrating azimuthally in the range
PA=6-10\degr\ and PA=34-38\degr\ showing the arcs (vertical tick
marks). Question marks indicate tentative arcs. }
\label{cutbeam}
\end{figure}

%% >>> searchlight beams and arcs.
Our K$_{\rm p}$ image has revealed two pairs of radial ``searchlight
beams'' emerging from the nebula center and directed oppositely
through the nebular center along PA$\approx$36\degr\ and
PA$\approx$8\degr\ (Fig.\,\ref{f2}).  The angular width of the beams
in the azimuthal direction is $\Delta$PA$\approx$15\degr.  The
searchlight beams are not axially symmetric, the beams running along
PA=36\degr\ being the brightest and the longest (observed up to
$\pm$2\farcs8 from the nebula center).  The north searchlight beam
directed along PA=8\degr\ is much fainter than its southern
counterpart. The south beams seem to be the continuation of the inner
and brighter $\wedge$-like feature observed in this lobe. There are
several arc-like features superimposed on the searchlight beams (see
also Fig.\,\ref{cutbeam}). Some of these arcs are also observed with a
much smaller brightness in the regions of the halo between the
beams. The separation between the arcs is in the range 0\farcs18 to
0\farcs24 and some of them appear to intersect each other. The
characteristic width of the arcs is $\la$0\farcs1.
%{\bf Raghvendra: NEED radial cuts through the searchlight
%beams to better characterize the width, separation, and brightness
%contrast of the arcs}.

The point where the beams cross at the nebula center (hereafter, beam
center) does not coincide with the center of the halo
(Fig.\,\ref{f2}). The latter has been determined by fitting elliptical
isophotes\footnote{Using the IRAF task
stsdas.analysis.isophote.ellipse} to the halo in K$_{\rm p}$ band: we
have only used the isophotes with intermediate brightness, making sure
that the main derived properties (center and shape) are not strongly
affected by the brightness distribution of the lobes or bright field
stars. The center of the beams has been estimated by taking azimuthal
cuts with different center points along the polar axis.  The center of
the beams is defined to give the largest degree of symmetry in the
width and position of the beams in the azimuthal cuts (we are assuming
that the beams are intrinsically symmetric). The point where the beams
cross at the nebula center and the center of the halo are separated by
53\,mas along the direction of the lobes, which is larger than the
expected errors in the location of the centers (errors
$<$\,1\,pixel$\sim$9mas).  Partial illumination of the halo by the
light scattered in the lobes (in addition to the starlight) could
explain the offset between the two centers and the similar orientation
of the halo to that of the lobes: the center of the halo is expected
to move in the direction of the brightest lobe, i.e.\, the NE lobe.

\subsection{Photometry}

\begin{figure}
\epsscale{0.5}
\plotone{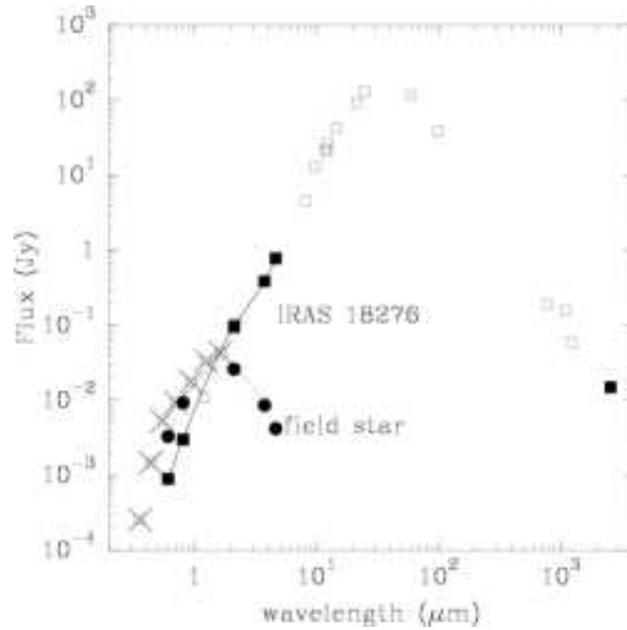}
\caption{Spectral energy distribution (SED) of \irase\ (squares) and the bright field star FS1 (circles; 
see Fig. 2). Data from this paper are plotted using filled
symbols. Literature data-points are represented by open squares (from the 2MASS and IRAS 
databases, Le Bertre et al. 1987, van der Veen et al. 1995, Meixner et
al. 1999). Crossed out symbols represent photometric data obtained
with large apertures most likely including flux from FS1 and that overestimate the flux from 
\irase. A solid
(dashed) line has been used to delineate the SED of \irase\ (FS1) in
the 0.6-3.8\micron\ range.}
\label{fig:sed}
\end{figure}

The optical and NIR photometry (and other observational details) are
given in Table \ref{tabphot}.  Total fluxes have been estimated by
integrating the surface brightness of all pixels within a polygonal
area enclosing the emission from our source with signal 3$\sigma$
above the sky and leaving outside bright field objects. Our fluxes in
the visible are significantly smaller than those obtained from
ground-based observations previously reported in the literature (see
Fig.\,\ref{fig:sed}).  This is because of the large apertures used in
ground-based observations, which most likely included flux from the
bright star FS1 north-west of \irase\ (Fig$.$\,\ref{f2}). Our flux
measurements in the NIR are also smaller than the 2MASS fluxes but are
in excellent agreement with the $J$ and $K$ photometry by Gledhill
(2005) obtained using an elliptical aperture including only flux from
\irase.
%% For reference we also give the  fluxes of FS1. K5 spectral type? 
%% Kp=26 mJy, Lp=8.6mJy, Ms=4.2 mJy. 
%% ### 2mass fluxes of the field star.
The integrated flux in the NE lobe is larger than in the SE lobe by a
factor 3-4 in the optical and 1.3-1.4 in the NIR (Table \ref{tabphot}).
%The dark waist has been chosen to be the boundary
%between the NE and SW lobe used to estimate the flux contribution from
%each lobe.

\subsection{Color maps}
\label{colormaps}

\begin{figure}
\epsscale{0.4}
\begin{center}
\rotatebox{270}{\plotone{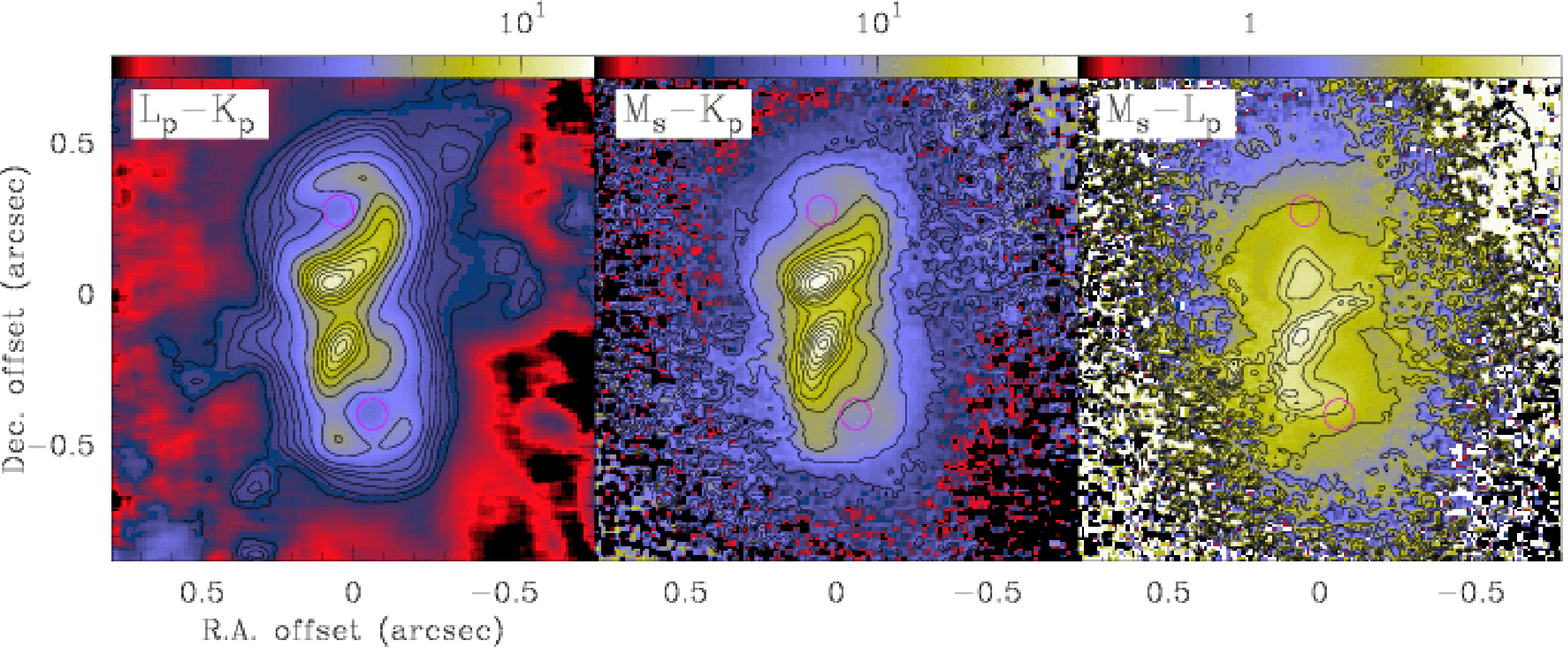}}
\epsscale{0.4} \rotatebox{270}{\plotone{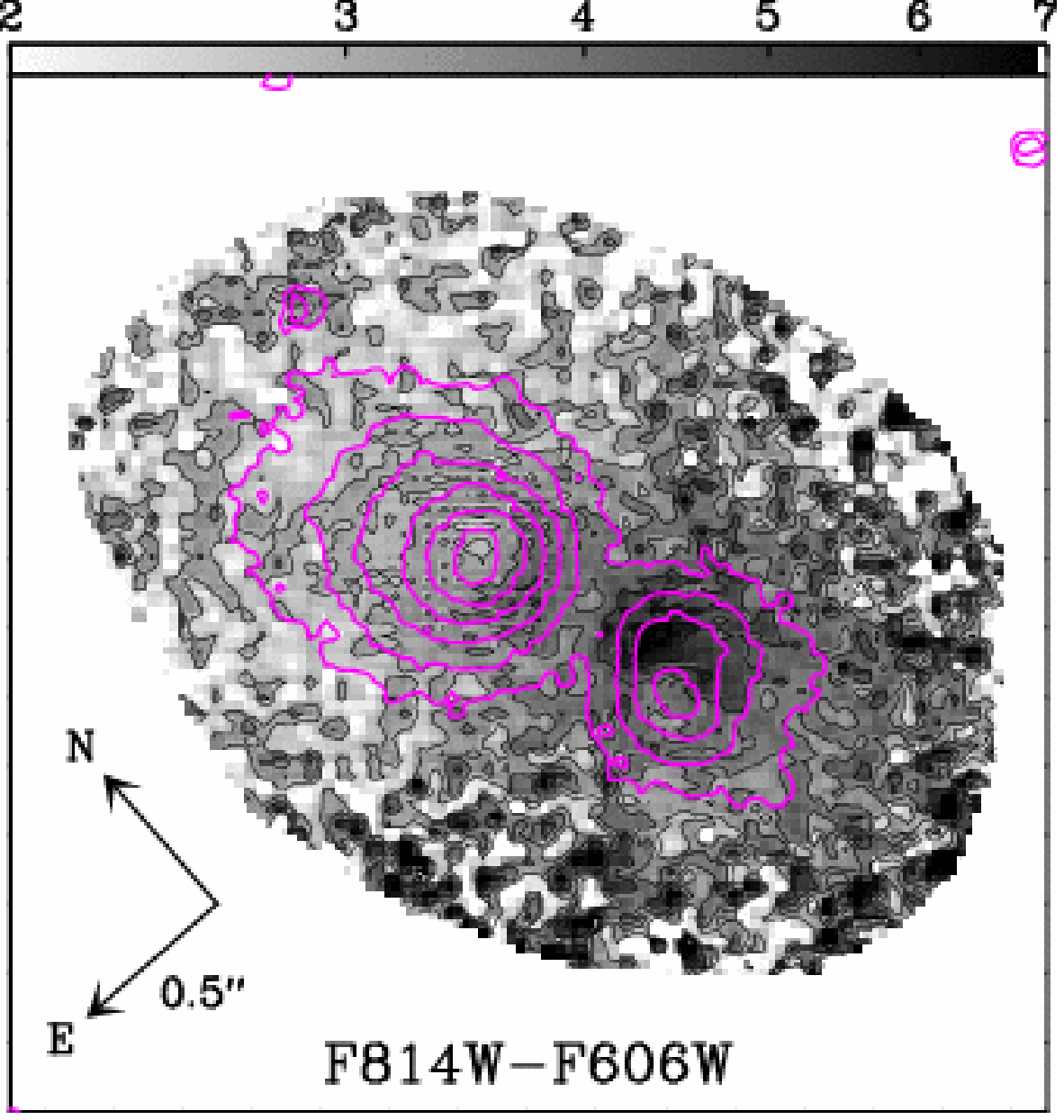}}
\end{center}
\caption{{\it Top)} Color maps obtained with the \kp, \lp, and \ms\ AO images. 
The color scale (in a logarithmic strecht) is shown in the wedge at
the top of the boxes.  Circles indicate the position and size of the
small, relatively blue patches used for studying the central
illuminating source (\S\,\ref{illum}). {\it Bottom)} \hstcol\ color
map (grey-scale); the F814W image (contours) is superimposed.}
\label{color}
\end{figure}

%In the case of a primary scattering nebula, the color maps are
%proportional to the column density and, therefore, are expected to be
%a good probe of the nebular structure.  
We have obtained color images
of \irase\ (F814W-F606W, \lp$-$\kp, \ms$-$\kp, and \ms$-$\lp), which are
shown in Figure\,\ref{color}.
%Optical-NIR color maps are not
%presented given the poorer quality of and significantly less
%structural detail provided by the $HST$ images compared with the AO
%images.
%% To match the two sets of images spatial smoothing, resampling, and 
%% rotation of the AO images would be needed, with the subsequent 
%% degradation of the images. 
%Registration of the $HST$ optical images has been done using field stars to
%compute relative object shifts\footnote{Using the IRAF task
%images.immatch.imalign}. Registration of the AO NIR images was done in two
%steps. {\bf First, a global registration was done by .... DAVID could you explain that?}). Second,
%we used the brightest field stars to compute small, residual object
%shifts (as for the $HST$ images). 
Registration of the $HST$ optical and NIR images has been done using field stars to
compute relative object shifts\footnote{Using the IRAF task
images.immatch.imalign}. The estimated uncertainty in the
image alignment is $\sim$1 pixel (i.e$.$, $\sim$10 and $\sim$23\,mas for
the AO and $HST$ images, respectively).

%The SW lobe is the reddest in our \hstcol\ color map .
The reddest region in our \hstcol\ color map is not in the waist but
in the base of the SW lobe, in a small region north of the relative
maximum in the F814W and F606W surface brightness distribution. There
is a color relative minimum at the waist, which is bluer than the base
of both lobes.

In our higher-resolution, better S/N ratio NIR color maps, the nebular
structure is much more clearly delineated.  The \lp$-$\kp\ and
\ms$-$\kp\ color images show an hourglass structure with an overall
point-like symmetry. The lobe walls and caps are quite well delineated
in these images, in particular, in the \lp$-$\kp\ maps. The West
(East) side of the North (South) lobe appears brighter. There is a
relative color minimum (i.e. a relatively blue region) in the middle
of the lobes, at a radial distance of 0\farcs3 from the center.  As in
the \hstcol\ color images, the \lp$-$\kp\ and \ms$-$\kp\ color images
show a relative minimum in the waist. The elongation of the halo along
PA=105-110\degr\ observed in the \lp-band images, is also noticeable
in the \lp$-$\kp\ image with relatively blue colors.
%% average color of 2.7 --- compared with the colors of the lobes. 

There are gradually increasing differences in the maps in going from
\lp$-$\kp\ to \ms$-$\kp\ and to \ms$-$\lp: the lobe walls and caps smooth out and
the waist reddens (relative to the lobes). In the \ms$-$\lp\ map the
color reddens gradually from the lobe tips to the waist, where the
nebula is reddest. These trends are mainly the result of the
lower scattering and absorption optical depths at these long
wavelengths. The waist in the \ms-\lp\ image shows a similar
orientation to the more extended equatorial structure seen in the
\lp-\kp, and \lp-band images.

%% ANYTHING ELSE TO MENTION HERE?? 

\subsection{Millimeter-wave observations}
\label{resultsco}

\begin{figure}
\epsscale{0.8}
\rotatebox{270}{\plotone{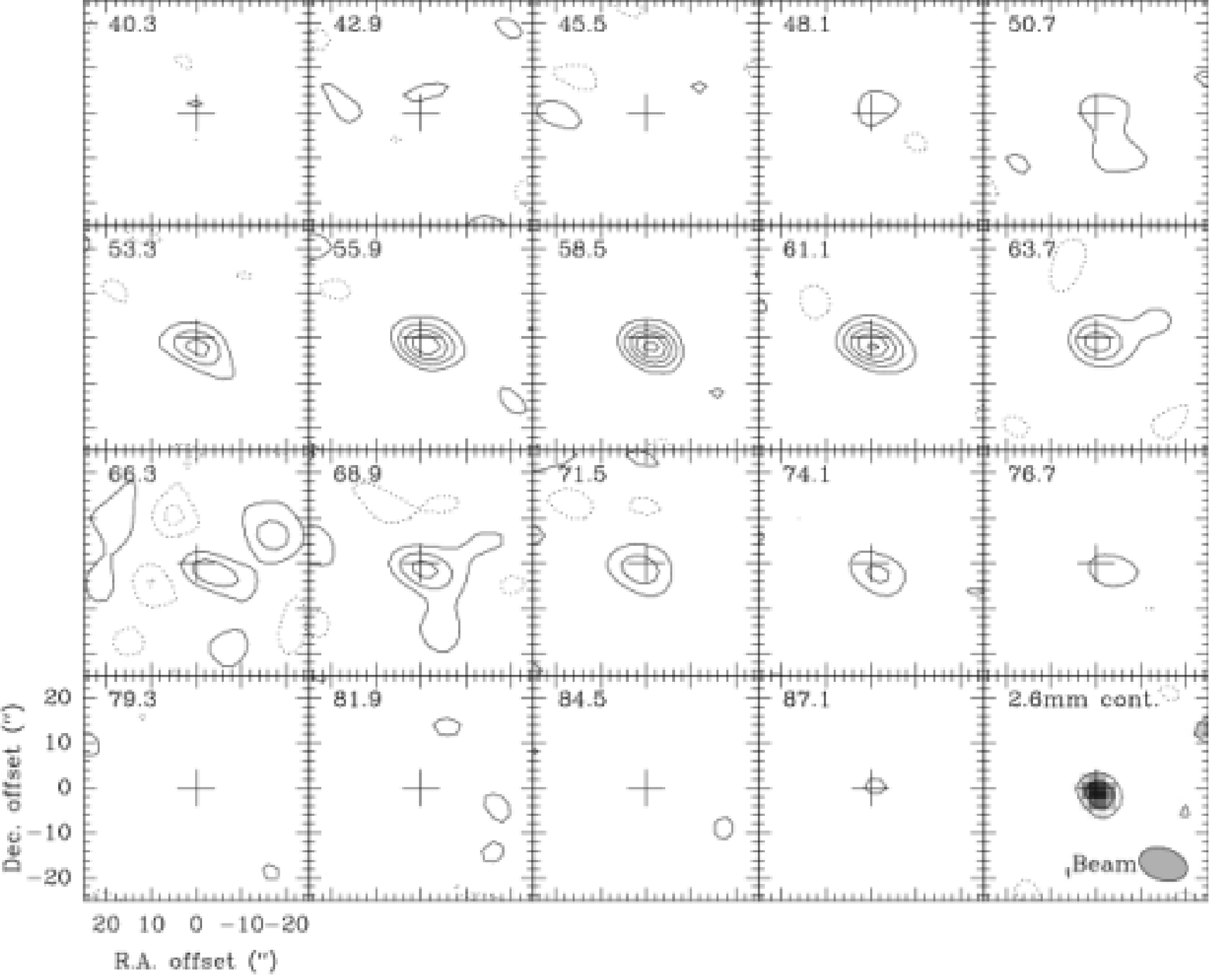}}
\caption{\doce\,1--0 velocity-channel maps of \irase\ obtained with OVRO. The LSR velocities 
of the different channels are indicated in \kms\ units within the
boxes. Levels are $-$4$\sigma$, $-$2$\sigma$, 2$\sigma$,
4$\sigma$... ($\sigma$=13Jy\,beam$^{-1}$). Positive (negative)
contours are plotted using solid (dashed) lines.  Brightness to
temperature units conversion factor is 1.19\,K per Jy\,beam$^{-1}$.
The 2.6\,mm continuum map and the clean beam,
FWHM$\sim$\,10\farcs9$\times$7\farcs1, are represented in the last
panel. Levels of the continuum are $-$2$\sigma$, 2$\sigma$, and
3$\sigma$ ($\sigma$=4mJy\,beam$^{-1}$). The coordinates of the center
of the maps (marked with a cross) are RA=18\h30\m30\fs75,
Dec=$-$14\degr28\arcmin56\farcs28 (J2000.0) }
\label{coone}
\end{figure}

\begin{figure}
\epsscale{0.32}
\plotone{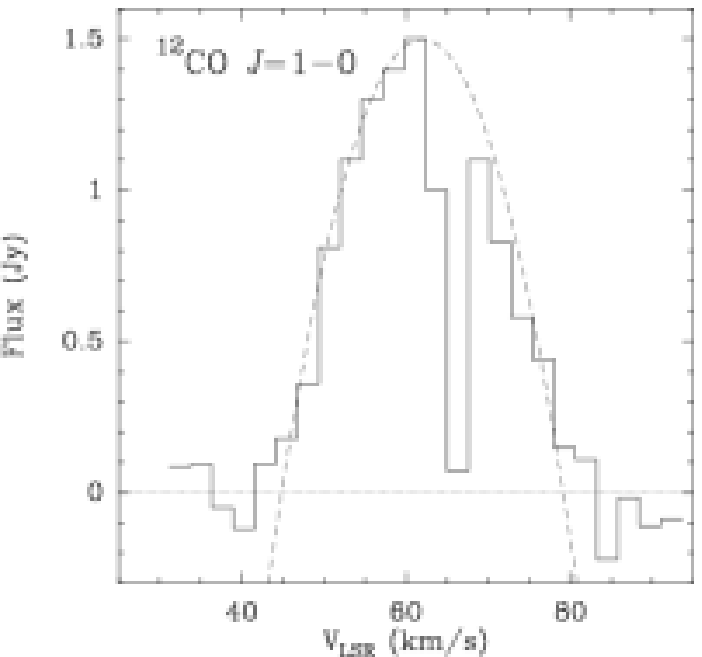}\plotone{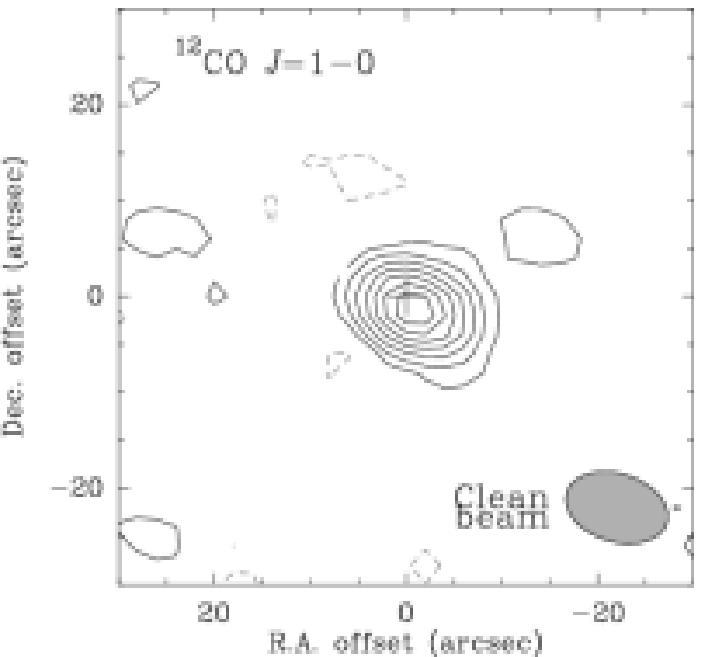}\plotone{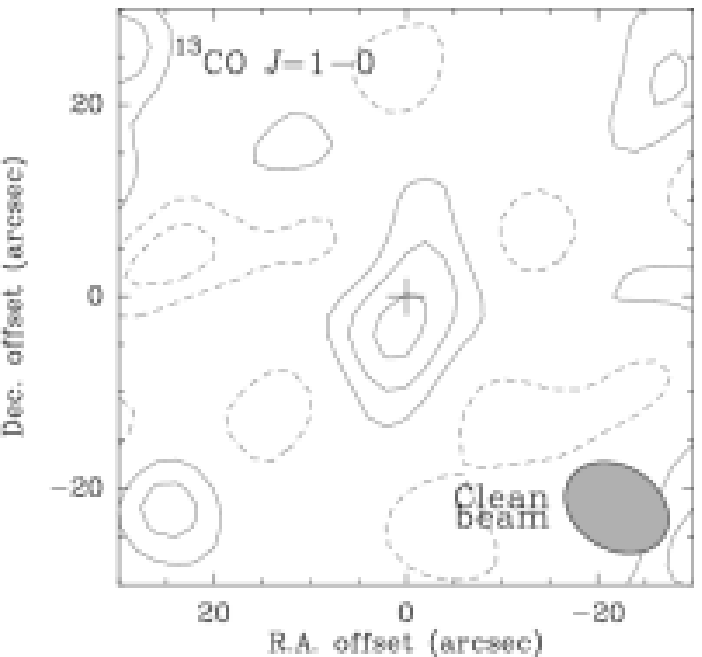}

\caption{$Left)$ \doce\,1--0 spectrum integrated over the nebula (solid line) and parabolic fit to the \doce\ profile (dashed line).  \doce\ 
  ($Middle$) and \trece\ ($Right$) intensity maps integrated over the
  line profile, i.e., within the LSR velocity range $\sim$[40:80\kms]
  and [53:59]\kms, respectively.  For \doce\ (\trece) levels are
  $-$2$\sigma$, 2$\sigma$,...  spaced every 2$\sigma$=3.4Jy\kms
  ($-$2$\sigma$, $-1$$\sigma$, 1$\sigma$ ... spaced every
  1$\sigma$=0.6Jy\kms.)}
\label{cotwo}
\end{figure}

Unresolved \doce\,$J$=1--0 emission  is observed towards \irase\ within
the LSR velocity range [+45:80]\,\kms\ (Figs.\,\ref{coone} and
\ref{cotwo}). The line profile is well fitted by a parabola centered
at \vsys=62\kms\ (LSR) and total width (at zero intensity level) of
34\,\kms. The parabolic profile is characteristic of unresolved
optically thick CO emission and yields an expansion velocity of the molecular
envelope of \irase\ of \vexp=17\kms. This value is comparable with,
but slightly larger than, that derived from previous single-dish
\doce$J$=2--1 spectra and OH maser observations (\vexp=12-14\kms; 
Heske et al., 1990; Bowers, Johnston \& Spencer, 1983).  
The observed \doce\ profile has a prominent, narrow absorption feature
in the range [+64:+69]\,\kms (LSR), i.e., bluewards of \vsys, which is
most likely due to interstellar absorption. For that
reason, the integrated flux over the line profile is expected to be a
lower limit to the intrinsic \doce\ flux from
\irase. A better estimate of the intrinsic flux from the source is given by 
the integral of the parabolic fit to the line profile, which we have used 
for the determination of the molecular mass in the envelope of 
\irase\ (Table \ref{tco} and \S\,\ref{comass}). The
\doce\ emission integrated over the line profile is
%offset from the center by 
%$-$1\farcs1, $-$1\farcs3 ($\pm$0\farcs6) from the center 
%(i.e., is 
centered at RA=18\h30\m30\fs67 Dec=$-$14\degr28\arcmin57\farcs6
(J2000.0), with an error of $\sim$0\farcs8 both in RA and Dec.

%Weak \doce\ emission wings could be present in the observed profile at LSR
%velocities of up to 80 and 40\,\kms, where the observed emission is
%larger than that given by the parabolic fit of the line core, but
%observations with higher S/N are needed to confirm this tentative result.
%%% molecular cloud identified?? 

\trece$J$=1--0 emission is detected above a 2$\sigma$ level in channels [+53:+58]\,\kms\ (LSR). 
Given its location, which coincides with the \doce\ and continuum
source, the \trece\ line most likely arises in
\irase\ but the observed profile is affected by strong interstellar absorption 
at LSR velocities [+64:+69]\,\kms\ or even a broader velocity
range, preventing us from obtaining an accurate characterization of
the line profile.  As for \doce, the measured \trece\ line flux is
most likely a lower limit to the total flux emitted by \irase.
Parameters derived from our \doce\ and \trece\ maps are given in
Table\,\ref{tco}.

An unresolved, continuum source is detected towards \irase\ with a
flux of F$_{\rm 2.6mm}$=15$\pm$4\,mJy.  The contamination of the
continuum flux by $^{12}$CO\,1--0 line emission is less than 5\%.

\section{Extinction towards the central star}
\label{extinction}

The extinction along the line of sight towards the central star in
\irase\ must be quite high not only in the visible but also at 
the observed infrared wavelengths since the illuminating star remains
hidden even in the \ms\ band. We have estimated a lower limit to the
 optical depth ($\tau$) at
2.12, 3.76 and 4.67\micron\ from the ratio between the observed
surface brightness ($I_{\rm obs}$) and the expected unabsorbed
intensity ($I$) per pixel from the central star.
%, which are related by
%I$_{\rm obs}$=I$e^{-\tau}$.  
Adopting values for the stellar effective temperature and luminosity
of \teff=7000\,K and $L$=6.4$\times$10$^3$\ls, a distance to \irase\
of $D$=3\,kpc (\S\,\ref{intro}), and taking into account the PSF in
our images, we derive: $I_{2.12\mu m}$=4.3$\times$10$^{-16}$,
$I_{3.76\mu m}$=9.9$\times$10$^{-17}$, and $I_{4.67\mu
m}$=1.7$\times$10$^{-17}$ erg\,s$^{-1}$\,\cm2\,\AA$^{-1}$ per pixel.
The non-detection of the star at our 3\,$\sigma$ level
(Table\,\ref{tabphot}) implies that the total (circumstellar plus
interstellar) extinction towards the nebula center along the line of
sight is $\tau_{2.12\mu m}$$>$12.6, $\tau_{3.76\mu m}$$>$8.6, and
$\tau_{4.67\mu m}$$>$5.4.  We have derived the interstellar extinction
towards \irase, $A_V$=1.6$\pm$0.5\,mag, using the numerical algorithm
provided by Hakkila et al$.$ (1997).  For a standard
total-to-selective absorption ratio R$_V$=3.1, the above value of
$A_V$ yields the following estimates of the interstellar extinction:
A$_{2.2\mu m}$=0.3\,mag, A$_{3.76\mu m}$=0.16\,mag, and A$_{4.67\mu
m}$=0.1.  Adopting these values, the $circumstellar$ optical depth
towards the star in \irase\ is $\tau_{2.12}$$>$12.3,
%$-$5log(D/3kpc)+2.5log($L$/3$\times$10$^3$\ls),
$\tau_{3.76}$$>$8.4, and $\tau_{4.67}$$>$5.3. 
%% this is also a lower limit because multiple scattering will enhance the intensity
%% close to the central star and will tend to connect the two lobes.
These values are in agreement with the order of magnitude of the
equatorial optical depth in the $J$ band, $\tau_J$$\approx$10,
inferred from modeling of the polarimetric data
\citep{gled05}, and the mean optical depth in the 10-25\micron\ range,
$\tau_{10-25\mu m}$$\ga$1, obtained from SED modeling (van der Veen et
al., 1995; see also \S\,\ref{sec:sed}), which indicates that the
equatorial envelope around \irase\ is optically thick.  Since the
optical depth in the $V$-band (0.55\micron) is typically a factor
$\sim$5 larger than in $K$-band (2.12\micron) (for silicate dust
$A_{\lambda}
\propto \lambda^{-1}$, see \S\,\ref{sec:sed}) the optical extinction
towards the star must be $A_V$$>$60\,mag. This value is well above
the lower limit to $A_V$ derived directly from our $HST$ F606W and
F814W images following the procedure described above,
$\tau_{0.606\micron}$$>$15.4 and $\tau_{0.814\micron}$$>$14.1.

\section{The illuminating source at the core of \irase}
\label{illum}
In this section we analyse the NIR colors of \irase\ and show that the
lobes are illuminated partially by the starlight and partially
by NIR light from a warm dust component, as has been also inferred
for the bipolar PPN IRAS\,16342-3814 (Sahai et al. 2005). The presence
of warm dust around the central star of \irase\ is also deduced from
the analysis of its full SED (Bedijn 1987 and \S\,\ref{sec:sed}).

We analyse the observed NIR colors averaged over small apertures (of
$\diameter$=0\farcs1) located in the middle of the lobes, where a
relative minimum in \lp$-$\kp\ is observed.  These regions (marked
with circles in Fig.\,\ref{color}) are well-removed from the lobe
edges and the dense central waist and, therefore, the optical depths to
scattered light are expected to be smallest.
%% Although \irase\ is known to have H_2 emission (K. Lowe, private comm.)
%% Measurements -----------------------------------------
%     L/K         M/K       M/L 
%NW   3.8         7.8       2.0  ** extinction corrected 
%SE   3.7         8.5       2.3  ** extinction corrected 
%% Measurements -----------------------------------------
In the NW- (SE-) lobe we find that the interstellar extinction
corrected intensities ($I_{\nu}$) for 2.1, 3.8 and 4.7\micron\ are in
the ratio 1:3.8:7.8 (1:3.7:8.5) indicative of a very red source
illuminating the lobes (note that the two lobes give very similar
colors). The observed colors in these regions result from a
combination of reddening due to internal extinction, and blueing due
to scattering, of light from the central illumination source.
Although H$_2$ emission has been detected in \irase, line emission is
not expected to be a significant contribution to the emission in the
NIR filters (Lowe et al., private communication --- see also the
featureless spectra of this object in the $H$, $K$, and $L$ bands
reported by Le Bertre et al., 1989).  Therefore, the observed
intensity of the scattered continuum in the reflection lobes
(corrected for interstellar extinction as in Sect. \ref{extinction}),
$I$, is given by:

\begin{equation}
\label{eq1}
%I \sim\ I_0 e^{-\tau_{\rm r}} e^{-\tau_{\rm t}} (1-e^{-\tau_{\rm los}^{\rm s}}) 
I \sim\ I_0 e^{-\tau}  (1-e^{-\tau_{\rm los}^{\rm sc}}) 
\end{equation}

\noindent
%where, $I_0$ is the emission from the central source, $\tau_r$ and
%$\tau_t$ are the extinction (absorption plus scattering) optical
%depths from the central source to the scattering region (radial) and
%from the scattering region to the observer (tangential), respectively,
%and $\tau_{\rm los}^{\rm s}$ is the line of sight scattering optical
%depth at the observed region. 
where, $I_0$ is the emission from the central source, $\tau$ is the
extinction (absorption+scattering) optical depth from the
central source to the scattering region and from the scattering region
to the observer (i.e$.$ radial+tangential), and $\tau_{\rm
los}^{\rm sc}$ is the line of sight scattering optical depth at the
observed region. 
%The optical depth $\tau_{\rm los}^{\rm s}$ accounts
%for scattering in the direction of the line of sight and, therefore,
%is expected to be a small fraction of the tangential scattering
%optical depth (which accounts for scattering in all directions
%different from the los). 
The above expression is valid for scattering
occurring predominantly in the lobe walls, which is a reasonable
approximation for \irase\ (\S\,\ref{lobes}). 

%The latter accounts for scattering in all directions different
%from los, and is expected to be smaller but comparable to the
%tangential absorption optical depth in the 2-5\micron\ range.  However,
%$\tau_{\rm los}^{\rm s}$ accounts for scattering in the direction
%of the line of sight, which is expected to be of the order of
%$\omega/4\pi$ for isotropic scattering, where $\omega$ is the solid
%angle subtended by the scattering region from the illuminating star.
%According to this, $\tau_{\rm los}^{\rm s}$ is expected to be $<$1
%even if the tangential absorption is $>$1.

We have modeled the \kp-\lp\ and \ms$-$\lp\ colors (allowing $\pm$20\%
errors) adopting the above formula and varying the temperature of the
illuminating source. The absorption and scattering optical depths have
been assumed to follow a power-law dependence with wavelength of the
type $\sim$$\lambda^{-1}$ and $\sim$$\lambda^{-4}$, respectively, in
the 2-5\micron\ range, as expected for silicate dust (e.g. Draine \&
Lee 1984; Suh 1999).  We find that the NIR colors measured cannot be
reproduced if the only illuminating source is the central star, which
is expected to have an effective temperature in the range
\teff=4000-10000\,K.
%For a $\lambda^{-2}$ dust emissivity, the NIR colors
%are even more difficult to reproduce. 
To explain the NIR colors, a warm (300-500\,K)
dust component is needed.
%% o maybe una lambda^-alfa  muy rara!! 
In particular, the best fit of the NIR colors is obtained if the
nebula is illuminated partially by the central star (for which we
adopted \teff=7000\,K) and partially by dust at 400\,K.  Our best fit
model indicates that a significant fraction ($\sim$80\%) of the
central star's light is absorbed by this warm dust cloud, which
presumably covers a significant solid angle around the star.  As shown
in \S\,\ref{sedmod}, 400\,K dust is expected to be located at
7-9$\times$10$^{14}$cm (45-60 AU) from the central star.  Our best fit NIR
color model also indicates that the light from both the star and the
dust cloud is extinguished by $\tau_{2.12\micron}$$\sim$3.7 in its way
to, and through the lobes.

%This value of the extinction, which is an upper limit to the radial
%extinction in the polar direction, is smaller that the extinction
%along the l.o.s. towards the central star (i.e. produced by the
%equtorial waist), implying that the pole-to-equator density contrast
%in \irase\ is $>$12.1/3.7$>$3.

\section{Structural components in \irase}
\label{analysis}

\subsection{The bipolar lobes}
\label{lobes}

\begin{figure}
\epsscale{0.35}
\rotatebox{270}{\plotone{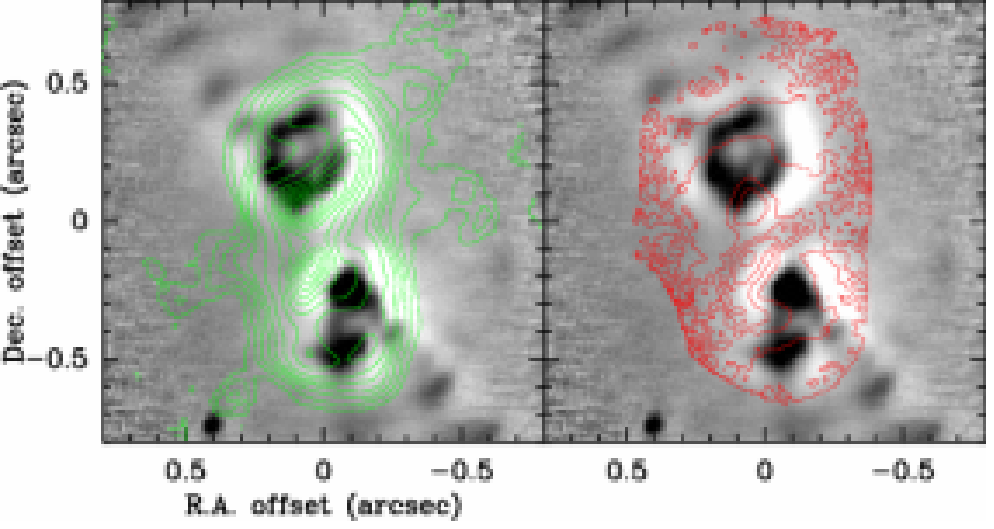}}
\rotatebox{270}{\plotone{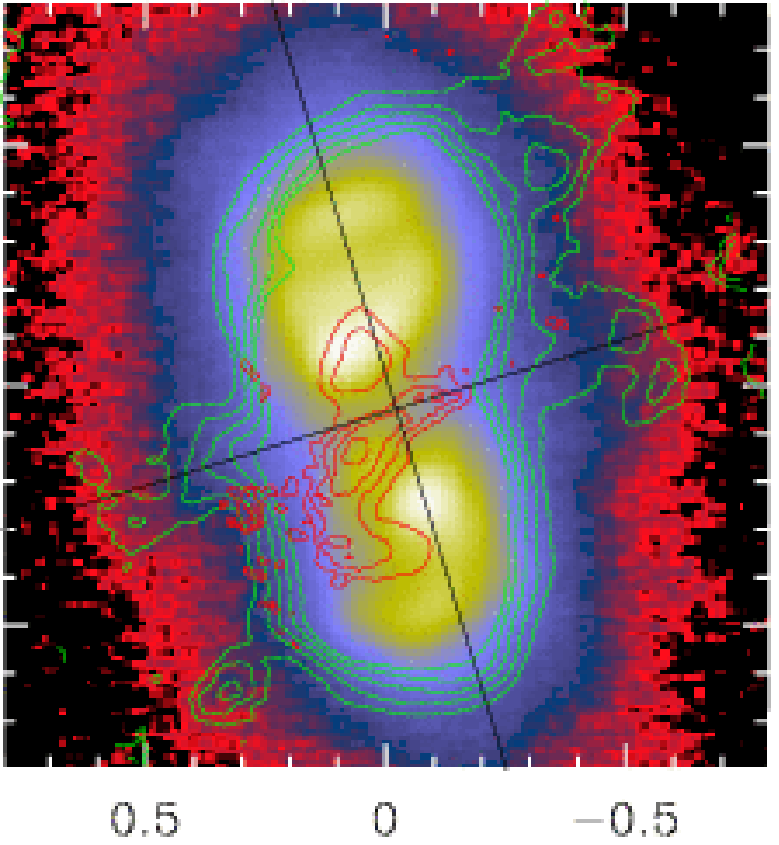}}
\caption{\lp$-$\kp\ ($left$) and \ms$-$\lp\ ($middle$) color maps
(contours) superimposed on the sharpened \kp-band image (grey-scale), and the
\lp-band image ($right$). In the right panel only the lowest (highest)
contours of the \lp$-$\kp\ (\ms$-$\lp) map are shown. The cross
plotted is located at the point where the beams cross at the nebula
center and is oriented along at PA=17\degr. Note the similar
orientation of the outer and inner equatorial structure probed by the
\lp$-$\kp\ and \ms$-$\lp\ color maps, and the \lp-band images.}
\label{use}
\end{figure}

Our NIR color maps and images suggest that the lobes of \irase\ are
most likely cavities excavated in the AGB envelope (i.e., structures
with dense walls and more tenuous interiors), the light from the
central source being reflected by dust in the lobe walls. 
%Previous scattering models of \irase\ indicate that the bipolar lobes
%are indeed regions of relatively low density (\S\,\ref{intro}). 
This is also believed to be the typical situation in many post-AGB
nebulae and, in particular for \irase, the OH maser spatio-kinematic
distribution suggests the presence of bipolar outflows carving out the
polar regions of the envelope (\S\,\ref{intro}).
Our \lp$-$\kp\ color maps indicate that the lobes are wider than they
appear in the \kp\ image and display a certain degree of point-symmetry
(see Fig.\,\ref{use}). Such a point-symmetry indicates that the
east and west walls of the lobes have different densities, with the
west (east) wall of the NE (SW) lobe being the densest, and/or are
illuminated differently from the central star.
%The \lp$-$\kp\ and
%\ms$-$\kp\ intensity ratio decreases with the distance from the base, which
%display the reddest colors, to the tips. This may indicate that the
%density in the lobe walls decreases with the distance to the center
%(note that the reddest colors are observed at the base of the lobes),
%assuming that the width of the lobe walls does not change
%significantly with latitude.
%The point-symmetric \lp$-$\kp\ and
%\ms$-$\kp\ color maps suggest 
The searchlight beam oriented at PA=36\degr\ seems to go through the least
dense lobe wall. The orientation of the lobes on the plane of the sky
as measured in the NIR color maps (PA$\sim$15\degr) is different  than in
the \kp\ image (PA$\sim$20\degr). The polarization angle measured by
Gledhill (2005), PA=23\degr, is very similar to the average PA of the
searchlight beams revealed by our AO images, and is probably a better
representation of the direction along which the bipolar cavities were
excavated. The different orientation of the lobes in the color maps is
most likely the result of large optical depth through the west and
east sides of the NE and SW lobes, respectively.  Although the north
and south lobes appear quite different in the \kp\ image, they are
similar in the color maps, suggesting that the real structure of the
two lobes is not that different and the $\wedge$-like appearance of
the SW lobe is mainly an illumination effect (presumably the same that 
produces the searchlight beams, which are similarily oriented).

Scattering structures will in general tend to smooth out in images at
long wavelengths because the scattering efficiency decreases as
$\lambda^{-4}$. This trend is observed in our images. 
%: the lobe edges
%and caps are quite poorly delineated in the \ms-band.  
%%At short
%wavelengths the extinction of the light from the central star to the
%scattering regions and through them may be very large and this will
%also avoid tracing the structure of these regions. This is the reason
%why the optical images show a featureless brightness distribution
%(although this is also partially due to their lower angular resolution
%compared with our AO images). 
The limb-brightened appearance of the lobe tips or caps in our NIR
images, which are best seen in the
\lp-band image, suggests that the scattering optical depth through these
regions is $<$1. The caps are not observed in the optical images,
which is
%,
%which show a smoothly decreasing intensity with the radial distance to
%the center, 
consistent with the larger optical depths through the caps at shorter
wavelengths. The structured lobe caps smooth out in the \ms\ image
most likely because the radial and tangential scattering optical depth
through the caps is too low at 4.67\micron: most of the light from the
central source that reaches the inner edge of the caps escapes in the
polar direction and very little is scattered towards us. The radial
cuts of the surface brightness along the lobes in the
\kp\ and \lp\ images show that the caps move outward in going
from \kp\ to \lp. The position of the lobe caps are very similar in
the \ms\ and \lp\ images. This implies that the extinction in the
polar direction through the caps at 2.12\micron\ is larger than 1. In
the \lp-band the lower extinction ($<$1) enables the photons to
penetrate deeper the lobe caps. These values enable estimating a range for
the optical depth through the lobe caps, 4\,$<$$\tau_V$$<$\,7, adopting
an extinction power-law $\sim \lambda^{-1}$.  This yields a range of
(2-4)$\times$10$^7$\cm3\ for the mean density in the
$\sim$0\farcs1-wide caps, assuming a gas column density to optical
extinction ratio of N$_{\rm d}$/$A_V\sim$3$\times$10$^{22}$\cm2 per
magnitude, which is expected for silicate grains and a dust-to-gas
mass ratio of $\delta$=200 (\S\,\ref{sedmod})
%% Nd/Av~1.5e21 for ism graphite+silicate dust.
Taking into account the size of the lobes and adopting a similar
density throughout the lobe walls and caps, the total mass in the lobe
walls must be $\ga$10$^{-2}$\,\msun. The lower limit to the mass
arises in the expected decrease of the density with the radial
distance to the nebula center.

%% compression factor at the lobe caps. Total mass in the lobe caps ~ 4e-5\ms.
%% mass in the lobes is >3.5e-3\ms, the lower limit arises because the 
%% density at the base of the lobes could be larger than at the tips given 
%% the reddest colors of the former. 

The redder F814W-F606W colors of the SW lobe can be interpreted as an
indication that the SW lobe is the farthest of the two, the larger
reddening being the combined result of 1) the presence of more
dust in the halo and the extended equatorial disk
lying in front of the distant lobe, and 2) more efficient forward
scattering (which would make the closer lobe appear brighter and
bluer than the far one). The NE lobe is also the brightest in all the
observed bands, which also favors its being the closest
lobe. Note, however, that intrinsic differences in the dust density
between the two lobes would affect their relative colors and
brightness (e.g. S\'anchez Contreras, Gil de Paz, \& Sahai, 2004). 
%The similar SW to NE lobe intensity
%ratio suggests that the bipolar lobes lie close to the plane of the
%sky, which is consistent with the low value for the inclination
%$i$=20\degr\ derived from OH maser observations.

\subsection{The halo} 

\begin{figure}
\epsscale{1.}
\rotatebox{270}{\plottwo{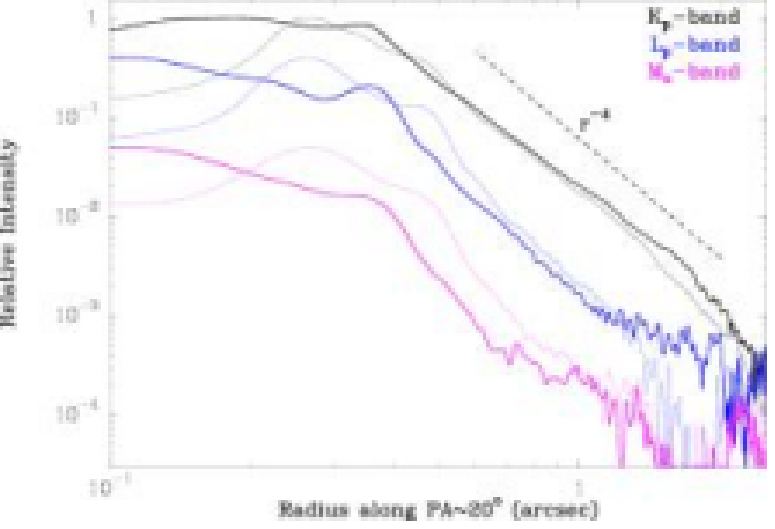}{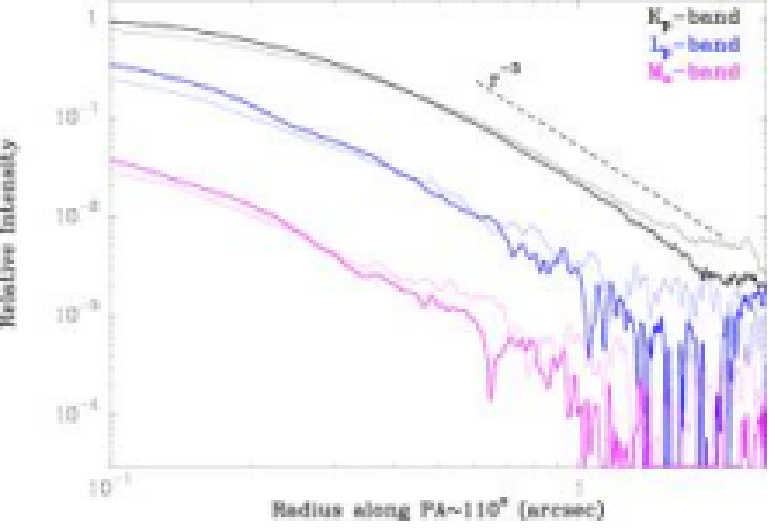}}
\caption{Surface brightness radial profiles 
through the center of the halo along the nebular major axis
(PA=15\degr-25\degr; $top$) and the equator (PA=105\degr-115\degr;
$bottom$) of the K$_{\rm p}$ (black), \lp\ (blue), and \ms\ (pink)
images.  Cuts are shifted by an arbitrary amount in the relative
intensity axis for clarity. Thick (thin) lines are used for the NE
(SW) lobe.  Radial power laws $\propto r^{-4}$ and $\propto r^{-3}$
are shown for comparison (dotted-lines). }
\label{cuthalo}
\end{figure}

The halo of \irase\ most likely represents part of the circumstellar
envelope ejected during the AGB phase that has not been strongly
affected by the lobe shaping processes. The halo is probably
spherical, the elongated shape being the result of the way the
different regions of the halo are illuminated, in particular, of the
lower optical depth along the poles and much larger extinction in the
equatorial direction. The outer radius of the halo is $\ga$2\farcs8,
as given by the length of the longest searchlight beams in the
\kp-band. The radial profile of the halo surface brightness
(Fig.\,\ref{cuthalo}) can give us information about the spatial
distribution of the density in the AGB envelope and how the mass-loss
rate that led to this component might have varied in the past.

A cut of the \kp-band image along the nebular symmetry axis shows that
beyond the tips of the lobes the halo radial profile varies with the
radial distance to the nebula center like $r^{-4}$. A similar variation
with $r$ is observed in the
\lp\ and \ms\ images. In the case of an optically thin
scattering halo, the surface brightness varies proportionally to the
light intensity dilution factor and to the line of sight scattering 
optical depth ($\tau_{\rm los}^{\rm sc}$) at each point. The light seen by the dust in the
halo along the polar direction (PA=20\degr) emanates mainly from the
central source, which is expected to be point-like. Therefore, the
light intensity dilution factor will vary as 1/$r^2$.  After
correcting the observed flux distribution by this factor, we deduce
that $\tau_{\rm los}^{\rm sc} \propto r^{-2}$ in the halo. Since $\tau_{\rm
los}^{\rm sc} \propto r^{1-\alpha}$ for a density power law $\propto
r^{-\alpha}$, we conclude that $\alpha$=3 in the halo. This result
agrees with the dust density distribution $\propto$$r^{-2.6}$ deduced
from SED modeling of
\irase\ \citep{veen95}. The previous result is consistent with the
AGB mass-loss rate increasing with time, assuming that the expansion
velocity of the wind is time-invariant.

The NIR surface brightness radial profile along the equator shows a
$r^{-3}$ dependence for radii larger than 0\farcs6. This is consistent
with the flux from these outer halo regions being mainly due to
second (or higher) order scattering. In fact, these equatorial regions
are mainly illuminated by the two bright scattering lobes rather than
by direct starlight, given the very large optical depth along the
equatorial direction ($A_V >$\,60\,mag). In this case, the light
intensity dilution factor is expected to have a dependence with $r$
shallower than that for a point-like illuminating source, probably of
the type 1/$r$, which is the dilution of a flux field emanating from
an infinitely long wire (see Alcolea al.\ 2000). After correcting by a
$\sim$1/$r$ dilution factor the observed flux distribution at radii
$\ga$0\farcs6, we obtain $\tau_{\rm los}$$\propto$$r^{-2}$ which is
consistent with a density law $\rho$$\propto$$r^{-3}$ as derived for
the polar regions.

The searchlight beams and arcs observed in the halo are most likely
illuminated parts of the AGB shell viewed in scattered
light. Discussion on the very uncertain origin of these features is
deferred to Section \ref{discuss}.

\subsection{The waist}
\label{waist}

The two lobes of \irase\ are separated by an equatorial dust
waist/torus behind which the central star is hidden.  OH maser
emission is observed to arise in this equatorial torus over a range of
angular radii of 0\farcs25 and 0\farcs85 (Bains et al., 2003,
\S\,\ref{intro}). The spatial coincidence between the NIR dark waist and the OH maser emitting region is expected since the OH masers are pumped by
FIR radiation from dust and are more likely to occur where the density
is highest. The faint, equatorial structure seen in the \lp-band image
with outer radius R$_{\rm out}\sim $0\farcs6 (\S\,\ref{resimaging})
probably corresponds to outer parts of the equatorial torus, which are
seen through multiply scattered light: a fraction of the photons
scattered in the lobes are directed towards the equatorial plane and
from there scattered again toward us. Although equatorial disks/tori
are usually inferred from the presence of dark lanes separating the
bipolar lobes, there are other PPNs were these equatorial structures
are observed in scattered light, e.g$.$ IRAS 04296+3429 (Sahai 1999),
IRAS 17106-3046 (Kwok et al. 2000), and IRAS 17347-3139 (S\'anchez
Contreras et al. 2006). The relatively blue colors of the waist in the
\lp$-$\kp\ and \ms$-$\kp\ color maps (\S\,\ref{colormaps}) are
consistent with multiple scattering illuminating with blue photons
these equatorial regions, which otherwise would be expected to show
the reddest colors.  Since scattering is much less efficient at long
wavelengths, the inner parts of the waist are best delineated in the
\ms$-$\lp\ color map, where the waist displays the reddest colors,
revealing the regions with the largest extinction along the line of
sight.

As shown in Section \ref{extinction}, the optical depth of the torus
towards the central star along the line of sight is
$A_V$$>$60\,mag. This value yields a column density of gas in front of
the star of N$_{\rm d}$$>$1.8$\times$10$^{24}$\cm2, for a gas column
density to optical extinction ratio of N$_{\rm
d}$/$A_V\sim$3$\times$10$^{22}$\cm2 per magnitude (see
\S\,6.1). 
Adopting an outer radius for the torus of R$_{\rm out}$=0\farcs85, as
given by OH maser emission mapping (see above), and assuming that the
inner radius of the torus is much smaller than R$_{\rm out}$ (as
indicated by SED modeling, \S\,\ref{sec:sed}), the mean H$_2$ density
in the torus
%(up to an outer radius of 0\farcs6=2.7$\times$10$^{16}$\,cm)
%(up to an outer radius of 0\farcs85=3.7$\times$10$^{16}$\,cm)
towards the central star along the line of sight is
$>$5$\times$10$^7$[$D$/3\,kpc]$^{-1}$\,\cm3.

%(see e.g. OH231, Alcolea et al. 2000).
%% This high value of the density is similar to that found 
%% equatorial waist of the well studied post-AGB nebula OH231.8 
%% (Alcolea et al. 2000). The total mass in the waist of OH231.8 is 0.64\ms.

An order of magnitude for the total mass in the waist enclosed within
a radius of $R$$\le$0\farcs85 and an height of
$h$$\le$0\farcs17\footnote{Projection effects can be ignored since the
lobes are very close to the plane of the sky and, therefore, the torus
is expected to be nearly edge-on (\S\,\ref{intro}).} can be roughly
estimated adopting a given density law as a function of the height
above the equatorial plane, $z$. In the following, we consider
two extreme cases for $\rho(z)$.  For a constant density throughout
the torus, $\rho(z)$=constant, the total mass in this component would
be $\ga$4\,[$D$/3\,kpc]$^{2}$\,\msun. A more realistic density law is
one in which the density decreases with increasing $z$
(\S\,\ref{resimaging}). For example, for a steep exponential density
law of the type $\rho(z)$=$\rho(0)\exp (-z\ln\chi/h)$ where
$\chi$=$\rho(0)$/$\rho(h)$, the mass in the torus would be
$\ga$4$\times$(1-1/$\chi$)/$\ln
\chi$\,[$D$/3\,kpc]$^{2}$\msun. Our NIR images suggest that the
optical depth through the waist at $h$ is probably $<$1 in the
2-4\micron\ wavelength range, since we start to see through the base
of the lobes (\S\,\ref{resimaging}).  Since the extinction in the
optical is typically a factor 5 larger than in the \kp-band, we
estimate that $A_V(h)$$<$5.  Therefore, a reasonable lower limit to
$\chi$ ($\propto$N$_{\rm d}(0)/N_{\rm
d}(h)$$\propto$$A_V(0)$/$A_V(h)$) may be $\approx$10, which yields a
total mass of $\approx$\,1\,[$D$/3\,kpc]$^{2}$\,\msun. The mass
derived is very uncertain since the total extent of, and density
distribution in, the torus as well as the geometry of the innermost
nebular regions, which determines the way the nebula is illuminated,
are unknown. Therefore the above figure has to be taken only as an
indication that the mass of this equatorial component could be quite
large, in particular, it could represent a significant fraction of the
total mass in the envelope. This is also observed in other post-AGB
nebulae, for example, in OH\,231.8+4.2, which has a dense
($\sim$3$\times$10$^6$\cm3) equatorial waist that contains $\sim$70\%
of the total envelope mass as derived from high-angular resolution CO
mapping (Bujarrabal et al., 2002; Alcolea et al., 2001).

For a silicate+graphite mixture in the envelope of \irase, which
cannot be completely ruled out (see \S\,\ref{sedmod}), the density and
the mass derived for the torus would be smaller. For example, the
standard ISM graphite and silicate mixture with a dust-to-gas mass
ratio $\delta$=200 yields N$_{\rm
d}$/$A_V\sim$1.5$\times$10$^{21}$\cm2 per magnitude. For this lower
value of the N$_{\rm d}$/$A_V$ ratio, the mean density in the torus
would be $\ga$2$\times$10$^6$\,[$D$/3\,kpc]$^{-1}$\cm3, leading to a
mass of $\sim$(0.2-0.07)\,[$D$/3\,kpc]$^{2}$\msun\ for the two extreme
(constant and exponential) density laws considered above.

%% This value is one order of magnitude larger than the lower limit to the 
%% molecular mass derived from the CO data. 

%Taking into account the outer radius of the torus, $R_{\rm
%out}$=2.7$\times$10$^{16}$cm, and assuming an expansion velocity of
%17\kms\ (see \S\,\ref{}), we estimate that the period of mass loss
%that led to this component started $\sim$500\,yr ago at a rate of
%$\approx$7$\times$10$^{-4}$ \my.

%It is worth mentioning the similarity of the equatorial waist of
%\irase\ to the equatorial torus in another well studied post-AGB
%nebula, OH231.8, which has a mean density of
%3.3$\times$10$^6$\cm3 and a total mass of 0.64\msun\ (Alcolea et al. 2000). 

%The
%kinematical ages of both torii are also similar (4500 yr for
%OH231). Since the linear size of the torus in OH231 is slightly larger
%than that in \irase, the mass in the former, 0.64\msun, seems a
%reasonable upper limit to the mass of the torus of \irase, assuming
%the same density spatial distribution in both cases.
%% WE COULD JUST SCALE FROM THE 
%% DIMENSIONS OF THE TORUS OH231.8

%Comparison with the OH maser envelope. 

\section{Mass, mass-loss, and dust temperature}
\label{phys}

\subsection{SED modeling}
\label{sec:sed}
We have derived the physical parameters of the dust envelope of
\irase\ by using a radiative transfer code to fit its SED at
wavelengths $\lambda$$\ga$5\micron, the region of the spectrum where
dust thermal emission is the main contributor to the observed flux
(Bedijn 1987). At $\lambda$$\la$2\micron\ the dominant contribution to
the flux is the scattered light from the central source. In the range
2 to 5\micron, a combination of these two emission mechanisms is
expected. Calculations were performed with the photoionization code
CLOUDY (version 05.07) last described by Ferland et al. (1998).
%Recent upgrades of the code are described in Ferland
%(2000a,b) and van Hoof, Martin, \& Ferland (2000). 
CLOUDY has a dust code built in that solves the time-independent
radiative transfer equation under the constraint of radiative
equilibrium. Details of the grain model are given in van Hoof et
al. (2004). The code makes a comprehensive, detailed treatment of the
grain physics that includes, e.g$.$, quantum and stochastic heating,
heating of the grains by Ly$\alpha$ photons, and the photo-electric effect.
%The importance and effects of physical
%processes involving grains can only be judged by including all these
%processes in a self-consistent manner.
CLOUDY does not work under the optically thin approximation, but
calculates the thermal dust continuum for optically thick envelopes.

We have assumed that the envelope is spherically symmetric.  Despite
the simplicity of this assumption, the code is expected to provide
%% reasonable results since we are interested in obtaining rough
reasonable constraints to the mean properties of the dust envelope (mass
and temperature). As shown by e.g.\ Ueta \& Meixner (2003),
asymmetries in the density distribution would mostly affect the
optical part of the SED but not the MIR/FIR, which is the spectral
region we are modeling. Scattering is assumed to be negligible in our
model.  As discussed in detail by Bedijn (1987), this is a reasonable
approximation at $\lambda$$\ga$2\micron\ for dust shells with
intermediate optical depths (i.e$.$ $\tau_{9.7\mu m}$$\sim$4), which
is the case of \irase. Even for very optically thick envelopes, with
$\tau_{9.7\mu m}$$\sim$30, inclusion of scattering does not change
significantly the SED in the wavelength range that we are modeling.

%% Asumptions and input parameters.
In order to calculate the SED of \irase\ we have adopted a series of
input parameters.  We have assumed that the central star has a
blackbody spectrum with an effective temperature of
\teff=7000\,K and bolometric luminosity of $L$=6400\,\ls (\S\,\ref{intro}). 
We have adopted astronomical silicate grains as suggested by the
presence of a weak 9.7\micron\ silicate feature in the IR spectrum of \irase\ 
\cite{LeBertre89,bed87}. 
%%  (MgSiFeO$_4$); dens_dust=3.3 gr/cm3.
%%  efficiency as that given in Martin and Rouleau 1991.
The optical constants for astronomical silicates are from the
calculations of Martin and Rouleau (1990), 
%%\citep{mar90}, 
which extend the work of Draine and Lee (1984)
%\citep{draine84}
to ionizing energies where the grains are
strongly absorbing. The high IR polarization measured in \irase\
suggests a population of small dust grains, with radii $a$$<$0.3\micron\,
\cite{gled05}. Therefore, we have used the grain size distribution
adopted by these authors, that is, with minimum and maximum grain
radii of $a_{\rm min}$=0.01\micron\ and $a_{\rm max}$=0.3\micron\, and
a power law size index of $-$5.5. CLOUDY uses a spherical Mie code to
generate the grain opacity as a function of wavelength for a given
size distribution and grain material (van Hoof et al., 2004).
%The extinction cross section of the dust in our model
%is The dust mass extinction coefficient (per unit dust mass) at
%100\micron\ used in our model is 24\cm2\,g$^{-1}$.
%The extinction cross section of the dust used in our model
%is as given in \cite{bald91}.
%

The other input parameters needed by the code to calculate the SED are
the inner and outer radius of the envelope, and the density
distribution.  For the density, we have adopted a radial power-law
$\rho \propto r^{-\alpha}$, and a dust-to-gas mass ratio constant
throughout the envelope with a value of $\delta$=200, which is typical of
post-AGB O-rich envelopes. As already suggested by Bedijn (1987), a
two-shell structure with warm and cool dust is needed to reproduce the
SED in \irase.
%The more massive, cooler outer shell would represent 
%the envelope resulted from heavy mass-loss during the AGB. 
The inner, warm dust shell would represent a recent low mass-loss rate
wind that started after the cessation of the heavy mass-loss at the
end of the AGB. The former period of heavy mass-loss led to the outer,
cooler envelope. In the outer shell, we have adopted $\alpha$=3, as
indicated by the radial profile of the halo in our NIR images. For
the inner shell we have used $\alpha$=2, which is expected for a
time-invariant mass-loss rate at a constant expansion velocity.
%% ###### 
%% probar que pasa con alpha=1?? o alpha=3?? 
%% ###### 
The free parameters of our model are: the inner radius of the inner
shell, $R_{\rm in,1}$, the inner radius of the outer shell, $R_{\rm
  in,2}$ (which is also the outer radius of the inner shell), the
outer radius of the outer shell, $R_{\rm out}$, and the density (or
equivalently mass-loss rate for \vexp=17\kms) at the inner
radius of each shell. These parameters have been varied until a good
match of the observations is obtained. The best-fit parameters have
been determined calculating and minimizing a $\chi^2$ function:

\begin{equation}
\chi^2=(F_{\lambda,\rm obs}-F_{\lambda})^2/\sigma_\lambda^2 
\end{equation}

where $\sigma_\lambda$ is the error of the observed flux
$F_{\lambda,\rm obs}$, which has been assumed to be of 20\% for all
wavelengths in the range modeled (i.e.\ 5-100\micron).

The parameters describing the dust envelope for the best-fit model are
given in Table\,\ref{tabsed}. The resulting synthetic SED together
with the radial dust temperature distribution are plotted in
Fig\,\ref{sedmod}. In this figure we also show the gas density
distribution for the best-fit model for a constant gas-to-dust mass ratio
throughout the envelope (see above) and the kinetic temperature of the
molecular gas predicted by CLOUDY after solving
the full physics (heating and cooling processes) 
for the derived envelope parameters.  

The total mass in the envelope of \irase\ enclosed within an outer
radius of $R_{\rm out}$=1.25$\times$10$^{17}$cm is $\sim$0.95\msun,
most of it is in the outer, cooler shell. Larger values of $R_{\rm
out}$ reproduce equally well the SED of \irase, therefore, the value
0.95\msun\ may be a lower limit to the total mass. (For an outer
radius of 10$^{18}$cm the total mass of the envelope would be
1.9\msun.) Although the best-fit SED model is obtained for an inner
radius of $R_{\rm in,1}$=6.3$\times$10$^{14}$cm, values in the range
$R_{\rm in,1}$=[5-8]$\times$10$^{14}$cm lead to acceptable results. The
optical depth is $A_V$$\sim$4\,mag, which is in good agreement with
the optical depth of the 9.7\micron\ silicate feature derived by
Bedijn (1987): note that the opacity of silicate grains as a function
of wavelength is such that $A_V$$\sim$$\tau_{9.7\micron}$.  The
extinction derived from SED modeling is much smaller than the
extinction towards the central star along the line of sight
($>$60\,mag) derived from our NIR images. This is expected since SED
modeling is done assuming spherical symmetry and, therefore, only
provides $average$ values for the envelope parameters.  The relatively
low value of $A_V$ derived from SED modeling compared with
$A_V$$>$60\,mag indicates a large equator-to-pole density constrast,
consistent with the optical and NIR images.  The dust temperature in
the inner shell can be approximated by a radial power-law of the type
$T_{\rm dust}(r) \propto r^{-0.4}$, except for the innermost regions
where the temperature increases with a steeper slope.
%from $r$$\sim$10$^{15}$\,cm to $R_{\rm in,2}$=1.6$\times$10$^{16}$.
In the outer shell, the dust temperature decreases 
outwards following a slightly shallower radial power-law, $T_{\rm dust}(r)
\propto r^{-0.34}$. Average gas densities in the inner and outer shell are in the range
$\approx$10$^5$-10$^8$\cm3 and $\approx$10$^7$-10$^4$\cm3,
respectively. Densities could be a factor $>$15 larger in the equator
and significantly smaller along the polar direction.

%%% ################
%%% COMMENT ON Rout 
%%% ################

%In our opinion, 
%grain composition is the main source of uncertainty in our model,
%given the tentative detection of the 9.7\micron\ silicate feature.

%% other outputs from CLOUDY: the equilibrium grain temperature ... 

%MENTION THAT THE MEAN OPTICAL DEPTH DERIVED FROM CLOUDY FOR THE 400K-SHELL 
%IS CONSISTENT  WITH THE EXTINCTION DERIVED FROM THE ANALYSIS OF THE NIR COLORS.

%% other things CLOUDY can provide: Tkin=H2!!  useful for CO?? 

Our model clearly underestimates the observed fluxes in the
submillimeter and millimeter wavelength ranges. Adopting a larger
value for $R_{\rm out}$ has little effect on the flux at long
wavelengths given the steep density law in the outer shell, therefore,
a larger $R_{\rm out}$ cannot explain the emission excess. The
emission excess could be partially accounted for by a dust emissivity
function shallower than that assumed in our model, $Q_{\lambda}
\propto \lambda^{-2}$, at wavelengths $\ga$100\micron. SED modeling of
a number of AGB and post-AGB objects in the submm and mm range suggest
that the absorption efficiency of the dust in O-rich envelopes follows
indeed a $\lambda^{-0.8}$ to $\lambda^{-1}$ law \citep{kna93,kna94}.
%, suggesting that the dust has amorphous or layered structure.  
However, from theoretical studies and laboratory works, an exponent
$\le$1.25 is only justified with difficulties \citep[][and references
therein]{gue96}. The submm and mm flux excess can also be explained if
there is an additional component of larger grains, as it has been
inferred for a number of post-AGB objects (e.g$.$ Walmsley et al.
1991, Jura et al. 1987, S\'anchez Contreras et al.\ 1998, Sahai et al.
2006).  The SED of \irase\ in the 800\micron-2.6mm range is very well
fitted by that of a black-body in the Rayleigh-Jeans limit, i.e$.$
$f_{\nu} \propto \nu^2$ (Fig.\,\ref{sedmod}).  Since grains are
expected to radiate like blackbodies at wavelengths
$\lambda$$<$2$\pi$$a$, an additional population of grains with radii
$a$$\ga$\,400\micron\ would explain the observed spectral index of the
continuum at long wavelengths.  The temperature of this
large-grain-component must be in the range $\sim$150-20\,K: if $T_{\rm
dust}$$>$150\,K then we would obtain too much flux at
$\lambda$$\la$100\micron; also $T_{\rm d}$ must be $\ga$20\,K since
the spectral index in the 800\micron-2.6mm is consistent with
black-body emission in the R-J limit.  Since the dust model built in
CLOUDY does not allow in its current version grains larger than
5\micron, we have estimated the mass of large grains needed to account
for the mm continnuum flux as explained in S\'anchez Contreras et al.\
(1998).  We have adopted a mean grain density and radius of $\rho$=3
g\cm3\ and $a \ga$ 400\micron, respectively.  Assuming that the dust
emissivity at $\lambda$$<$2.6mm is $Q_{\lambda}$$\sim$1, we estimate
that a total mass of dust of $M_{\rm
dust}$$\ga$1.6$\times$10$^{-4}$\,[$D$/3\,kpc]$^2$\msun, for $T_{\rm
dust}$=150\,K, and $M_{\rm dust}$$\ga$
1.6$\times$10$^{-3}$\,[$D$/3\,kpc]$^2$\msun, for $T_{\rm dust}$=20\,K,
is contained in this big grain dust population.  Adopting a
gas-to-dust mass ratio of $\delta$=200\footnote{In the absence of
reliable estimates of $\delta$ for the large-grain component in
circumstellar envelopes, we adopt the same value as for the 'normal'
dust component.}, the total gas mass associated with this
large-grain-component is
$M$$\sim$(0.03-0.3)\,[$D$/3\,kpc]$^2$\,[$\delta$/200]\msun. The
location of this large-grain-component is unknown, since the
mm-continuum emission is unresolved in our $\sim$8\arcsec-resolution
maps.  Cold (20-30\,K), big grains have been inferred to be located in
the dusty equatorial regions of some PPNs, where the grains are
shielded against the starlight due to the large extinction along the
equator (e.g.\ IRAS\,22036+5306, Sahai et al., 2006).  The temperature
of the large-dust-component in \irase\ is not well constraint and,
therefore, there is not a sufficiently strong reason to claim that the
grains are in the dark (cold) waist. Big grains are also believed to
grow in long-lived, orbiting disks (e.g.\ Red Rectangle, Jura et al.\
1997).
%Relatively large grains are present in the lobes of some PPNs (S\'anchez
%Contreras et al., 1998). 
High-angular resolution mapping of the sub-mm
or mm continuum emission and/or measurements of the continuum in the
$\lambda$$\sim$300-500\micron\ range are needed to unveil
the location of the big grains in \irase.

Finally, our model predicts a 9.7\micron-silicate feature that is
slightly stronger than observed, however, we have not found any models
that simultaneously produce a weaker silicate feature and provide a
good fit of the SED. This may suggest that there is a certain
mixture of grains with different composition in \irase.

\subsection{The mass from CO data}
\label{comass}

\begin{figure}
\epsscale{0.5}
%%\rotatebox{270}{\plotone{cont9_dem.ps}}
\rotatebox{270}{\plotone{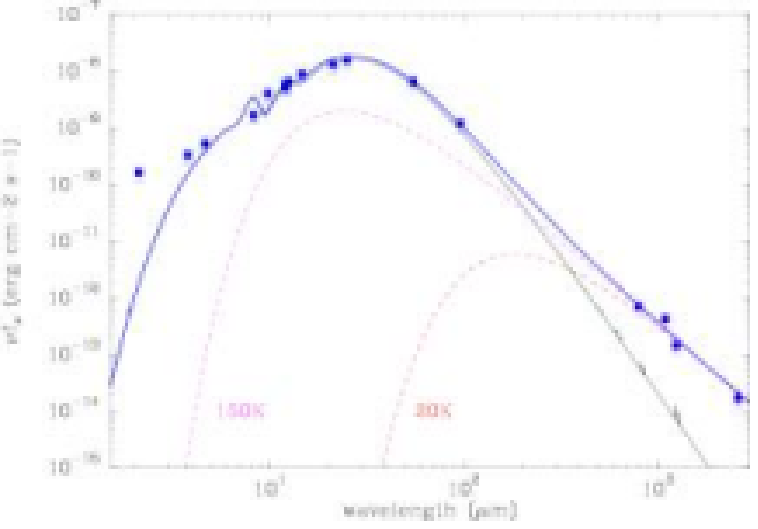}} 

\rotatebox{270}{\plotone{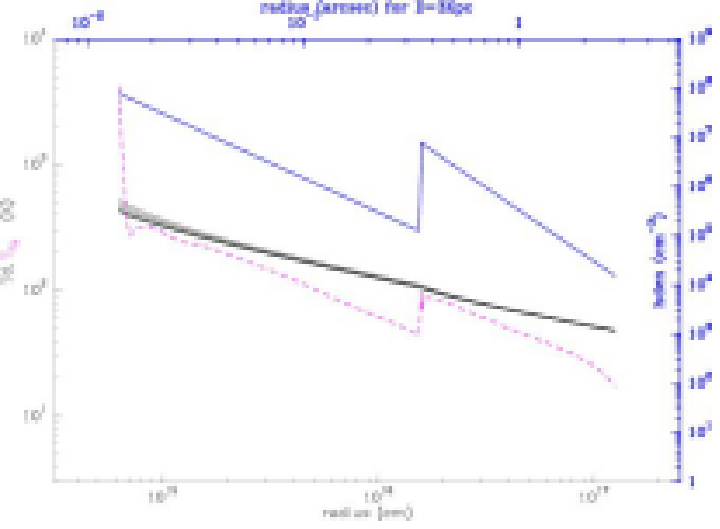}} 
\caption{$Top)$ Observed SED of \irase\ corrected by interstellar extinction (squares), SED predicted by CLOUDY for our best-fit model (thin-solid, black
  line), and blackbody spectra for $T_{\rm dust}$=150\,K and $T_{\rm
  dust}$=20\,K (dashed lines).  A combined model spectrum (best-fit
  CLOUDY model plus 150\,K black-body) is shown with the thick-solid,
  blue line. $Bottom)$ Spatial variation of the grain equilibrium
  temperature for the different grain-size bins used in the model
  (solid, black lines). The gas density (blue line), obtained by
  adopting a constant dust-to-gas mass ratio thoughtout the envelope
  ($\delta$=200), and the gas kinetic temperature (pink, dashed line)
  predicted by CLOUDY for the best-fit envelope parameters derived
  from SED modeling are also shown.}
\label{sedmod}
\end{figure}

As for most PPNs, the CO emission most likely arises in an expanding
envelope that is the remnant of the AGB envelope and that is probably
the counterpart to the NIR halo and waist. We have estimated the mass
of this component from the \trece\ and \doce\,1--0 flux. We have
assumed that both lines are thermalized, i.e$.$ the CO excitation is
described by a rotational temperature that it is similar to the
kinetic temperature, $T_{\rm kin}$. The thermalization assumption is a
reasonable approximation for most PPNs and, in particular, is expected
to be valid for \irase, which has nebular densities larger than the
CO\,$J$=1--0 critical density ($\approx$10$^3$\,cm$^{-3}$) in most
regions of the CO-emitting envelope probed by our maps ($R_{\rm
CO}$$<$5\arcsec) as shown in Fig.\,\ref{sedmod}.  We have used the
mean kinetic temperature of H$_2$ predicted by CLOUDY for our best-fit
model (Fig.\,\ref{sedmod}) to estimate the molecular mass from CO. We
have adopted a mean temperature of $T_{\rm kin}$=110\,K and a relative
\doce-to-H$_2$ abundance of 2$\times$10$^{-4}$ and
\trece-to-H$_2$ abundance of 10$^{-5}$, which are common values in
PPNs (Bujarrabal et al. 1997, 2001, and references therein), and the
well known molecular parameters of CO (partition function, Einstein
coefficients, etc; see e.g. Rohlfs and Wilson, 2000). We derive a
total mass of molecular gas in \irase\ of $>$0.28\,\msun\ from \doce\
(using the integrated flux of the parabolic fit to the line profile)
and $>$0.38\,\msun\ from \trece\ (Table \ref{tco}). The lower limit to
the mass derived from \doce\ is because this line is optically thick
as indicated by its parabolic profile. Additional confirmation is
given by the intensity ratio between the \doce\,$J$=2--1 line, which
was observed by Heske et al$.$ (1990) with the IRAM 30\,m
radiotelescope (beam=21\arcsec) with a main beam temperature of
$T_{\rm mb}$=1\,K, and the $J$=1--0 line, for which we measure $T_{\rm
mb}$=1.8\,K with a 10\farcs9$\times$7\farcs1 beam. This yields a
beam-corrected $J$=2$-$1-to-$J$=1$-$0 line ratio of $\sim$1, which
indicates that both lines are optically thick through most regions of
the CO-emitting envelope.  The lower limit to the mass derived from
the \trece\ line arises because of the strong interstellar absorption
of the circumstellar emission and maybe because of a large optical
depth in this transition as well. The fact that the mass derived from
\doce\ is smaller than that derived from
\trece, which is expected to be less opaque, is in agreement with the
\doce\,$J$=1--0 transition being optically thick.

The lower limit to the mass derived from CO is in agreement with the
total mass obtained from SED modeling, $\sim$1\,\msun.  The total
mass enclosed within the upper limit to the outer radius of the
CO-emitting nebula probed by our maps, i.e.  within 5\arcsec\
(2$\times$10$^{17}$cm), is 1.2\,\msun,
which is also consistent with the lower limit to the mass obtained
from CO.

We note that values of \tkin\ of the order of $\approx$10$^2$\,K are
expected and observed in young PPNs with dense envelopes (e.g.\,CRL
618, S\'anchez Contreras et al.\ 2004, see also Hrivnak \& Bieging
2005). Values of \tkin\ as lows as 15-30\,K are also observed,
however, in a number of PPNs (Bujarrabal et al., 2001). For a mean
value of the kinetic temperature of 50\,K, which is the average of
\tkin\ over volume for \irase\ (Fig.\,\ref{sedmod}), the lower limit
to the mass derived from \trece\ is $\sim$0.2\msun.

\section{Formation and evolution of \irase}
\label{discuss}

The analysis of our high-angular resolution NIR and optical images,
and the SED of \irase\ enable us to trace back in time the evolution
of this object and to attempt understanding how the different
structural components revealed by our data were formed. From the
values for the outer and inner radius of the outer dust shell of
\irase\ ($R_{\rm out}$$\ga$1.25$\times$10$^{17}$\,cm and $R_{\rm
  in,2}$=1.6$\times10^{16}$\,cm, Table\,\ref{tabsed}) and the
expansion velocity derived from CO, \vexp=17\kms, we infer that the
heavy mass-loss process that led to this component started at least
$\sim$2300\,yr ago and ceased $\sim$300\,yr ago. During that time the
mass-loss rate increased from \mloss$\sim$2$\times$10$^{-4}$\,\my\, to
2$\times$10$^{-3}$\my.  This heavy, time-increasing mass-loss-rate
wind, which is expected at the end of the AGB phase, was succeeded in
\irase\ by a lower mass-loss rate wind
(\mloss$\sim$3$\times$10$^{-5}$\,\my), which could have stopped, or
been replaced with an even more tenuous wind recently
($\approx$10-15\,yr ago).  The presence of bipolar lobes in \irase\
may imply large expansion velocities in some regions of the envelope
(mainly along the polar axis). Although most of the dust is expected
to be located in the massive AGB envelope, some dust fraction is
located in shock-accelerated regions (e.g., the lobe walls), which are
expected to have a complex and unknown kinematics.  Therefore, the
ages derived above assuming \vexp=17\kms\ are uncertain, especially
for the inner dust shell.
%Our images and SED modeling provide a lower limit to the outer radius,
%and kinematical age, of the AGB envelope of \irase\ (see above).
%The heavy mass-loss rate at the tip of the AGB phase is expected to last
%$\approx$10$^4$\,yr but very unlikely more than 10$^5$\,yr.  For an
%outer radius of the envelope $R_{\rm out}$$<$10$^{19}$\,cm (which
%corresponds to $<$10$^5$\,yr for \vexp=17\kms), the mass in the
%envelope predicted by our SED model is $\sim$3\msun.  This can be
%considered as an upper limit to the mass of the progenitor star.

As proposed for most PPNs (\S\,\ref{intro}), the lobes of \irase\ have
been probably excavated by interaction of fast, bipolar (jet-like)
winds with the AGB envelope. The jet-launching mechanisms which are
believed to operate in the early post-AGB or late AGB phase are still
the subject of a vigorous debate. From the deprojected total length of
the lobes (0\farcs5$\sim$2.2$\times10^{16}$\,cm) and adopting an
expansion velocity of $\approx$100\kms, which is typical of bipolar
post-AGB outflows (e.g$.$ Bujarrabal et al. 2001), we estimate a value
of $\sim$65\,yr for the kinematical age of the lobes. As shown by our
NIR color maps, the lobes display a certain degree of
point-symmetry. Such a symmetry is found in many post-AGB objects and
is often interpreted as a signature of precessing jet-like winds
(e.g. Sahai et al. 2005).
%This value is probably 
%In the case of a young object such as \irase, the previous value
%of \tkin\, is not expected to be 
%not a reliable indicator of the time
%elapsed since the beginning of the jet+'AGB envelope' interaction in this case, 
%since the acceleration and shaping of \irase\ are probably taking place at present

%% waist 
The origin of the equatorial waist/torus is very uncertain. Given the
large extent of the torus compared to the bipolar lobes, we believe
that the waist is unlikely the result from the interaction of fast
jets with an initially isotropic AGB wind. If the waist is just what
is left of the AGB CSE after the jet+`AGB envelope' interaction, then
we should observe a spherical envelope with a central biconical cavity
that is as large as the optical lobes. In contrast, the dense material
in the waist obscuring the central star and the base of the lobes is
constrained to the equatorial region forming a torus that extends up
to a radius of $\ga$0\farcs6 ($\ga$2.7$\times10^{16}$\,cm), as derived
from our \lp-band images, and maybe of $\ga$0\farcs85
($\ga$3.7$\times10^{16}$\,cm), as derived from OH maser emission
mapping (\S\,\ref{intro}). It may be possible for shocks developed in
the jet+`AGB envelope' interaction to move sideways and backwards
compressing the gas in the equatorial plane and shaping the
torus. However, since these shocks are expected to be much weaker and
slower than the forward shocks which produce the lobes, they can form,
at best, a torus whose radius is much smaller than the length of the
lobes, contrary to what we observe.  Alternatively, the waist of
\irase\ could be the result of a period of enhanced mass-loss concentrated in
the equatorial plane. However, what induces such a heavy and
asymmetrical mass-loss is unknown, especially considering that the
mass in the disk, although very poorly determined, could be a
significant fraction of the total mass in the envelope
(\S\,\ref{waist}). Adopting an expansion velocity of \vexp=17\kms, we
estimate that the kinematical age of the torus is $\ga$700\,yr.

%% searchlight beams 
Searchlight beams and arcs like those observed in \irase\ are two of
the most fascinating, yet poorly understood features revealed by
recent high-angular resolution, high sensitivity images of AGB and
post-AGB objects. Beams and arcs are most likely illuminated portions
of the envelope viewed in scattered light.  As proposed by Sahai et
al. (1998), searchlight beams may result from starlight escaping
through holes/cavities in an inner dust cocoon. The agent excavating
the holes is unknown but it may be the same collimated (precessing)
jets responsible for carving out the bipolar lobes.  Morris (1981) had
originally proposed another, more simple, scenario to explain the
structure of bipolar reflection nebulae including searchlight beams,
which were observed as small protrusion on the reflection lobes of
some PPNs in {\sl low-angular resolution} images. In this scenario,
which was also adopted in the scattering models by, e.g., Yusef-Zadeh,
Morris, \& White (1984) and Latter et al. (1993), these protrusions
result from a density law that declines with latitude above the
equatorial plane if, at some latitude, the radial optical depth to
starlight falls rapidly below unity. As extensively discussed by Sahai
et al. (1998), the latitude-dependent density model, however, was not
able to explain satisfactorily the detailed morphology of the
searchlight beams in the PPN CRL 2688 as observed with {\sl
high-angular resolution}.  Some of the objections to such model argued
by Sahai et al. (1998) also apply to \irase. In particular, our images
do not show any evidence of systematic ellipticity in the arcs as
would be expected in Morris' scenario.  Moreover, the
latitude-dependent density scenario does not account for the sharp
edges and relatively large width of the searchlight beams in \irase\
($\Delta$PA$\approx$15\degr;
\S\,\ref{resimaging}). 
Nor does it account for the point-symmetry of the searchlight beams as
observed in our \kp-band image (\S\,\ref{resimaging}), which would
require a point-symmetric density distribution at a spatial scale as
large as the extended halo, where the beams are observed up to a
radial distance of $\pm$2\farcs8. Such a density law would be very
difficuly to support since it would imply a point-symmetric mass loss
during the AGB phase.  As far as we know, searchlight beams are also
observed in the PPNs: CRL2688 \cite{sah98}, IRAS 20028$+$3910
\cite{hri01}, IRAS 17150$-$3224 \cite{su03}, and IRAS 17245$-$3951
\cite{hri99}, raising to five the number of PPNs with searchlight
beams. The increasing number of objects with beams suggest that these
are not extremely rare features, and, therefore, the dust distribution
required to explain the beams must be relatively common at least
during some time in the evolution of these objects. 
%In order to
%attempt understanding the origin of then beams, first, we must ask
%ourselves: What do all these 5 objects have in common? 
It is worth noting that the 5 objects listed above display bipolar
morphology, have dense, equatorial waists, and are seen nearly edge-on
(IRAS 20028+3910 appears to be at an intermediate orientation, Su et
al.\,2001). The presence of a nearly edge-on, dark waist may then be a
necessary condition for the observation of the beams. This is in part
expected because scattering features will be necessarily more
difficult to observe in objects with very bright cores outshining the nebula.
%% but maybe equatorial density enhancement is required to produce the beams.... 
Another interesting aspect is that beams are not observed in PNs or
AGB envelopes, but only in young PPNs.  The lack of beams in PNs may
be an observational effect, since their bright central stars are
usually directly visible and the scattering nebula is very
diluted. The lack of beams in AGB envelopes is consistent with the
idea that the winds clearing out the holes in the cocoon start after
the star has left the AGB. The spectral types of the searchlight
beam's PPNs are concentrated in the F-G range. This may be partially
due to destruction of the dust cocoon by the high-velocity outflow
responsible for carving out the holes and/or by the increasing stellar
UV field as the star evolves towards hotter spectral types. Finally,
the presence/absence of searchlight beams does not seem to be strongly
correlated with the chemistry, since beams are observed both in C- and
O-rich objects, however a larger sample of searchlight beam PPNs is
needed to reach reliable conclusions.

\begin{figure}
\epsscale{0.5}
\plotone{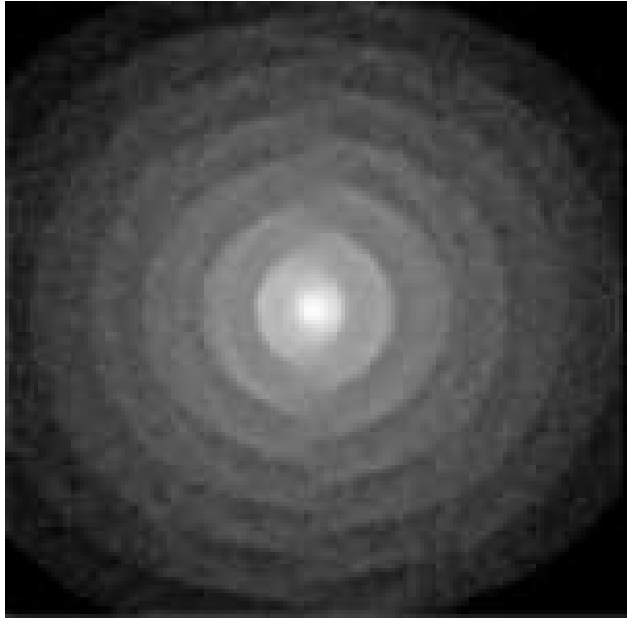}
\caption{Model of spiral density shells induced in a circumstellar  
outflow by orbital motion in a binary system.  The gray-scale shows  
the projected dust density for an observer lying in the system's  
orbital plane -- projected as a horizontal line running through the  
center.  (The projection from most other angles has a spiral form.)   
To compare with the observations of \irase, the dust distribution  
should be convolved with the illumination pattern, which is strongest  
along a cone centered on the system's angular momentum axis -- a  
vertical line through the center.  The details of this model are  
described separately (Morris et al., in preparation).}
\label{spiral}
\end{figure}

%% arcs... 
Arcs and rings are also getting more and more common in PPNs.  Up to
date arcs have been detected in 7 PPNs (including \irase), 4 PNs, and
1 AGB envelope (see summary of arcs/rings in post-AGB objects by Su
2004). 
The binary-induced spiral outflow model originally proposed by
Mastrodemo and Morris (1999) produces arc-like features very similar
to those observed in \irase\ for a system viewed from near the
equatorial plane and taking into account illumination effects (such as
very large extinction along the equator by the dusty waist).  In this model (see
Fig.\,\ref{spiral}), the arcs disappear along the polar axis and
alternate in radius on each side of the bipolar axis. The alternating
arcs spaced along the polar axis are best seen if the angular momentum
axis of the system is perpendicular to the line of sight. From our
point of view, the most attractive aspect of the binary-induced spiral
outflow model is the ability to explain the non-concentric,
intersecting arcs of \irase. This model has also been proved to
explain the spiral pattern recently discovered in the extended
envelope of the PN AFGL 3068 (Mauron \& Huggins, 2006; Morris et al.,
in preparation).  A possible argument against this model is that
predicts arcs separated at quite regular intervals, set by the orbital
period, whereas the time spacing of the arcs in \irase\ ranges between
190 and 250\,yr for \vexp=17\kms. Models involving magnetic activity
cycles (e.g.\ Soker 2000, Garc\ai a-Segura et al.\ 2001) appear as an
interesting alternative possibility to explain non-regularly spaced
arcs. However, it is not known whether or not magnetic periodic
variations can happen on a scale of hundreds of years, as required to
explain the arc spacing in most PPNs (Su 2004).
%In this model the orbital period sets
%the separation between the arcs, which suggests $P$$\sim$190-250\,yr
%for \irase. 
%All PPNs with searchlight beams show arcs (except maybe 17245, although)! 
%%maybe the process leading
%to both features are somehow related. That is clear, arcs are BEST seen in the beams in iras18276, due to the lower optical depth through the cavities.
%which the light escapes leading to the beams expected to dissapear
%with time?

%This object looks like iras17441 and 17150 - peanut-like ... 
%similarities? same type of object??

\section{Summary}
\label{summary}

We present and analyse NGS-AO images in the \kp, \lp, and \ms\ filters
of the PPN \irase\ obtained with the Keck\,II 10\,m telescope.
Supporting optical F606W and F814W $HST$ images as well as
interferometric observations of the \doce\,($J$=1--0),
\trece\,($J$=1--0), and 2.6\,mm continuum emission with OVRO are also
presented. Our main findings are summarized as follows:

\begin{itemize}
  
\item[-] Our AO images reveal the structure of the dust envelope
  around \irase\ with 0\farcs06-0\farcs1-resolution. The envelope of \irase\ 
  displays a clear bipolar morphology with two lobes separated by a
  dark waist and surrounded by a faint, sensitivity-limited halo.  Our
  \kp-band image reveals two pairs of radial ``searchlight beams''
  emerging from the nebula center and several intersecting, arc-like
  features.

\item[-] The lobes are probably cavities with dense walls and caps excavated
  in the AGB envelope presumably by tenuous bipolar winds. The
  point-symmetry displayed by the nebula may suggest that these
  bipolar winds are precessing. Although quite uncertain, our estimate
  of the total mass in the lobe walls is $\ga$10$^{-2}$\,\msun. For an
  adopted lobe expansion velocity of 100\,\kms, which is typical of
  post-AGB outflows, we estimate a kinematical age $\sim$65\,yr for
  this component.
  
\item[-] The halo most likely represents the remnant of the AGB CSE that
  has not been strongly affected by the process that produces and
  shapes the lobes. The density in the NIR halo follows a radial
  power-law $\propto$$r^{-3}$, which is consistent with a mass-loss
  rate increasing with time during the latest stages of the AGB. The
  outer radius of the halo is $\ge$1.25$\times$10$^{17}$cm, which
  implies that the halo was ejected $\ga$2300\,yr ago, for an
  expansion velocity of \vexp=17\,\kms\ (see below).

\item[-] The illuminating star remains hidden behind the dark waist
  even in the \ms-band images. Such a waist most likely represents a
  dense equatorial torus, also traced by OH maser emission, that could
  have resulted from equatorially enhanced mass loss.  The
  $circumstellar$ optical depth towards the star along the line of
  sight (i.e., through the torus) derived from our NIR images is
  $\tau_{2.12}$$>$12.3, $\tau_{3.76}$$>$8.4, and $\tau_{4.67}$$>$5.3,
  which implies a large extinction in the optical, $A_V$$>$60\,mag,
  and a mean density $>$5$\times$10$^7$\,\cm3. The mass in the
  equatorial torus is very uncertain, but it could represent a large
  fraction of the total mass of the envelope.

\item[-] The searchlight beams may result from a preferred
  illumination of the halo through ``holes'' excavated in an inner
  dust cocoon, maybe by the same agent responsible for carving out the
  bipolar lobes.
  
\item[-] The intersecting arcs, which are spaced at time intervals
  ranging between 190 and 250\,yr, are reasonably well explained by
  the binary-induced spiral-outflow model originally described by
  Mastrodemos \& Morris (1999), which produces arc-like features very
  similar to those observed in \irase\ for a system viewed nearly
  edge-on and taking into account illumination effects (Morris et al,
  in preparation).

\item[-] The analysis of the NIR colors indicates that the scattering
  nebula is partially illuminated by the central star (with
  \teff=4000-10,000\,K) and partially by a warm ($\sim$300-500\,K)
  dust component. The presence of warm dust is confirmed by modeling
  of its full SED.
  
\item[-] We have modeled the 5-100\micron\ wavelength range of \irase\ 
  using the dust radiative transfer model built in CLOUDY.  Based on
  the detection of a weak 9.7\micron-silicate feature, astronomical
  silcates have been used in our model. The grain size distribution
  used is the same as that adopted by Gledhill (2005) to explain
  NIR polarimetry data. The central star has been assumed to emit like
  a \teff=7000\,K, $L$=6400\ls\ black body at a distance $D$=3\,kpc.
  
\item[-] SED modeling suggests a two-shell structure: 1) an outer dust
  shell with inner and outer radius $R_{\rm
    in}$$\sim$1.6$\times$$10^{16}$cm and $R_{\rm
    out}$$\ga$1.25$\times$10$^{17}$cm, dust temperature $T_{\rm
    d}$$\sim$105-50\,K, and mass-loss rate that increased from
  \mloss$\sim$2$\times$10$^{-4}$\my\ to 2$\times$10$^{-3}$\my\ in
  $\sim$2000\,yr (adopting an expansion velocity \vexp=17\kms, see
  below); and 2) an inner shell with $R_{\rm
    in}$$\sim$6.3$\times$10$^{14}$\,cm, $T_{\rm dust}$$\sim$500-105\,K, and
  \mloss$\sim$3$\times$10$^{-5}$\my.
  
\item[-] The mass of the envelope enclosed within $R_{\rm
    out}$=1.25$\times$10$^{17}$cm derived from SED modeling is
  $\sim$1\msun, for a dust-to-gas mass ratio of $\delta$=200.

\item[-] The molecular envelope of \irase\ is unresolved in our
  $\sim$8\arcsec-resolution \doce\ maps. Our CO data indicate that the
  bulk of the CO-emitting nebula is expanding at low velocity,
  \vexp=17\,\kms. Adopting a mean kinetic temperature in the envelope
  of \tkin=110\,K, as predicted by CLOUDY for the best-fit model to
  its SED, we derive a mass of $M$$>$0.38\,\msun\ for the molecular
  envelope.
  
\item[-] Our best-fit SED model  underestimates the
  observed sub-mm and mm flux. This excess can be accounted for by an
  additional population of big grains (radius $a$$\ga$0.4\,mm) with
  temperature and mass in the range $T_{\rm dust}$=150-20\,K and $M_{\rm
  dust}$=(0.16-1.6)$\times$10$^{-3}$\msun, respectively. The location of
  this large-grain-component is unknown, since the 2.6mm continuum
  emission is unresolved in our maps. One possibility is that big
  grains are located in the dense equatorial waist.

\end{itemize}
\acknowledgments

The authors thank the anonymous referee for his/her critical comments
and valuable suggestions that have helped improving significantly this
paper.  The data presented herein were obtained at the W.M. Keck
Observatory, which is operated as a scientific partnership among the
California Institute of Technology, the University of California, and
NASA. The Observatory was made possible by the generous financial
support of the W.M. Keck Foundation. This work has been partially
performed at the California Institute of Technology and the Department
of Molecular and Infrared Astrophysics of the {\sl Instituto de
Estructura de la Materia, CSIC,} and has been partially supported by
National Science Foundation Grant No$.$\,9981546 to Owens Valley Radio
Observatory and the Spanish MCyT under project DGES/AYA2003-02785.  RS
is thankful for partial financial support for this work from a NASA/
ADP grant.  AGdP is financed by the MAGPOP EU Marie Curie Research
Training Network and partially by the Spanish {\sl Programa Nacional de
Astronom\'{\i}a y Astrof\'{\i}sica} under grant AYA2003-01676.  MM and
RS acknowledge financial support for this work from a NASA/LTSA
grant. This research has made use of the SIMBAD database, operated at
CDS, Strasbourg, France, and NASA's Astrophysics Data System.


%% in parentheses, after the keyword, but they will not be verified.

{\it Facilities:} \facility{Keck II (NIRC2)}, \facility{HST (WFPC2)}, \facility{OVRO}.



\clearpage 

\begin{table}
%\caption{1.6 - 4.7$\micron$ photometry}
\begin{center} 
\caption{AO \& HST observations \& photometry of \irase} 
\label{tabphot}
\begin{tabular}{lcccccccc} 
\tableline\tableline 
%% Delta lambda = filter width 
filter & $\lambda_{0}$ & $\Delta \lambda$ &time  &FWHM  & rms ($\sigma$)   & $I_{\rm max}$ & Flux & NE-to-SW   \\
       &  (\micron)    & (\micron)        &(sec) &(mas) & (mJy/pix)        & (mJy/pix)     & (mJy)& Flux ratio  \\
\tableline
F606W\tablenotemark{a}  & 0.61 & 0.123 & 200 &138                      & 4$\times$10$^{-6}$ & 3.8$\times$10$^{-3}$ &0.9 & 4    \\
F814W\tablenotemark{a}  & 0.81 & 0.149 & 120 &138                      & 1$\times$10$^{-5}$ & 1.1$\times$10$^{-2}$ &3.0 & 3    \\ 
K$_p$\tablenotemark{b}  & 2.12 & 0.351 & 270 &60--65\tablenotemark{c}  & 8$\times$10$^{-6}$ & 4.3$\times$10$^{-2}$ &96  & 1.3  \\
L$_p$\tablenotemark{b}  & 3.78 & 0.700 & 225 &80--85\tablenotemark{c}  & 3$\times$10$^{-4}$ & 0.29  		   &385 & 1.4 \\
M$_s$\tablenotemark{b}  & 4.67 & 0.241 & 180 &95--100\tablenotemark{c} & 2$\times$10$^{-3}$ & 0.70                 &781 & 1.4 \\
\tableline
\end{tabular}
\tablenotetext{a}{HST subsampled images: FoV=35\arcsec$\times$35\arcsec; 22.8\,mas\,pixel$^
{-1}$}  
\tablenotetext{b}{Keck AO images: FoV=10\arcsec$\times$10\arcsec; 9.94\,mas\,pixel$^{-1}$} 
\tablenotetext{c}{Range determined from PSF fitting to several field stars at the same radial distance from the reference star than our target.}
%% \tablecomments{\bf Any more comments here?} 
\end{center}
\end{table}

%%%%%%%%%%%%%%%%%%%%%%%%%%%%%%%%%%%%%%%%%%%%%%%%%%%%%%%%%%%%%%%%%%%%%%%
\begin{table}
\caption{CO line parameters and derived molecular mass}
\label{tco}
\begin{tabular}{lcc} 
\hline\hline 
& \doce\,1--0 & \trece\,1--0 \\
\hline
\vsys\ (\kms) \dotfill & 62 & \tablenotemark{a} \\
\vexp\ (\kms)\dotfill & 17 &  \tablenotemark{a} \\
$I_{\rm CO}$ (Jy\,beam$^{-1}$) \dotfill & 1.5$\pm$0.3 & 0.32$\pm$0.08 \\
$\int$ $I_{\rm CO}$ d$V$ d$\Omega$ (Jy\,\kms)\dotfill & 27$\pm$3
(34\tablenotemark{b}) & 2\tablenotemark{c}$\pm$1 \\
 & & \\
\hline
X(CO/H$_2$) \dotfill & 2$\times$10$^{-4}$ & 10$^{-5}$ \\
\tex\ (K) \dotfill & 110 & 110 \\
$M_{\rm mol}$ (\msun) \dotfill & $>$0.28\,$[D/3$kpc]$^2$ & $>$0.38\,$[D/3$kpc]$^2$ \\
\hline 
\end{tabular}
\tablenotetext{a}{A reliable estimate of this parameter is not possible due to low S/N}  
\tablenotetext{b}{Integrated flux of the parabolic fit to the \doce\ line profile (see Fig.\ref{cotwo} \& \S\,\ref{resultsco})}
\tablenotetext{c}{Affected by strong insterstellar absorption}
\end{table}
%%%%%%%%%%%%%%%%%%%%%%%%%%%%%%%%%%%%%%%%%%%%%%%%%%%%%%%%%%%%%%%%%%%%%%%

%%%%%%%%%%%%%%%%%%%%%%%%%%%%%%%%%%%%%%%%%%%%%%%%%%%%%%%%%%%%%%%%%%%%%%%
\begin{table}[htbp]
\caption{Envelope parameters from SED modeling}
%%\begin{center} 
\begin{tabular}{lr} 
\tableline\tableline 
\multicolumn{2}{c}{\it Input source parameters} \\
\teff(K) & 7000 \\
$L$(\ls) & 6400 \\
$D$(pc)  & 3000 \\
\multicolumn{2}{c}{\it Input grain parameters} \\
Composition & Astronomical silicates \\
$a_{\rm min}$(\micron) & 0.01 \\
$a_{\rm max}$(\micron) & 0.3  \\
n($a$) & $\propto a^{-5.5}$  \\
gas-to-dust mass ratio & $\delta$=200 \\
\tableline 
\multicolumn{2}{c}{\it Derived parameters} \\
\multicolumn{1}{c}{\underline{Inner shell}} \\
$R_{\rm in,1}$(cm)  & 6.3$\times10^{14}$\\
$R_{\rm out,1}$(cm) & 1.6$\times10^{16}$ \\
$T_{\rm d}$(K) & 500-105\\
$\rho(r)$ & $\propto r^{-2}$ \\
\mloss(\my) & 3.2$\times10^{-5}$\\
$M$(\msun) & 0.007 \\
\multicolumn{1}{c}{\underline{Outer shell}} \\
$R_{\rm in,2}$(cm) & 1.6$\times10^{16}$ \\
$R_{\rm out,2}$(cm) &  1.25$\times$10$^{17}$ \\
$T_{\rm d}$(K) & 105-50\\
$\rho(r)$ & $\propto r^{-3}$ \\
\mloss(\my) & 2.0$\times10^{-3}$\\
$M$(\msun)  & 0.95 \\  %% Mtotal=2.36 
\multicolumn{1}{c}{\underline{Large-grain-component}} \\
$a$ & $\ga$ 400\micron \\
$T_{\rm d}$(K) & 150-20\\
%$M_{\rm tot}$(\msun) & (0.16-1.6)$\times10^{-3}$ $\times$ $\delta$$\dagger$\\
$M$(\msun) & 0.03-0.3\\
\tableline\tableline 
\end{tabular}
%% \tablecomments{Masses and mass-loss rates} 
%\tablenotetext{\dagger}{Assuming the same gas-to-dust mass ratio for the big-grains component.}
%%\end{center}
\label{tabsed}
\end{table}
%%%%%%%%%%%%%%%%%%%%%%%%%%%%%%%%%%%%%%%%%%%%%%%%%%%%%%%%%%%%%%%%%%%%%%%

\clearpage
%%%%%%%%%%%%%%%%%%%%%%%%%%%%%%%%%%%%%%%%%%%%%%%%%%%%%%%%%%%%%%%%%%%%%%%
%%%%%%%%%%%%%%%%%%%%%%%%%%%%%%%%%%%%%%%%%%%%%%%%%%%%%%%%%%%%%%%%%%%%%%%

\clearpage 
%%%%%%%%%%%%%%%%%%%%%%%%%%%%%%%%%%%%%%%%%%%%%%%%%%%%%%%%%%%%%%%%%%%%%%%
%%%%%%%%%%%%%%%%%%%%%%%%%%%%%%%%%%%%%%%%%%%%%%%%%%%%%%%%%%%%%%%%%%%%%%%

\clearpage 
%%%%%%%%%%%%%%%%%%%%%%%%%%%%%%%%%%%%%%%%%%%%%%%%%%%%%%%%%%%%%%%%%%%%%%%
%%%%%%%%%%%%%%%%%%%%%%%%%%%%%%%%%%%%%%%%%%%%%%%%%%%%%%%%%%%%%%%%%%%%%%%

\clearpage 
%%%%%%%%%%%%%%%%%%%%%%%%%%%%%%%%%%%%%%%%%%%%%%%%%%%%%%%%%%%%%%%%%%%%%%%
%%%%%%%%%%%%%%%%%%%%%%%%%%%%%%%%%%%%%%%%%%%%%%%%%%%%%%%%%%%%%%%%%%%%%%%

\clearpage 
%%%%%%%%%%%%%%%%%%%%%%%%%%%%%%%%%%%%%%%%%%%%%%%%%%%%%%%%%%%%%%%%%%%%%%%
%%%%%%%%%%%%%%%%%%%%%%%%%%%%%%%%%%%%%%%%%%%%%%%%%%%%%%%%%%%%%%%%%%%%%%%

\clearpage 
%%%%%%%%%%%%%%%%%%%%%%%%%%%%%%%%%%%%%%%%%%%%%%%%%%%%%%%%%%%%%%%%%%%%%%%
%%%%%%%%%%%%%%%%%%%%%%%%%%%%%%%%%%%%%%%%%%%%%%%%%%%%%%%%%%%%%%%%%%%%%%%

\clearpage 
%%%%%%%%%%%%%%%%%%%%%%%%%%%%%%%%%%%%%%%%%%%%%%%%%%%%%%%%%%%%%%%%%%%%%%%
%%%%%%%%%%%%%%%%%%%%%%%%%%%%%%%%%%%%%%%%%%%%%%%%%%%%%%%%%%%%%%%%%%%%%%%

\clearpage 
%%%%%%%%%%%%%%%%%%%%%%%%%%%%%%%%%%%%%%%%%%%%%%%%%%%%%%%%%%%%%%%%%%%%%%%
%%%%%%%%%%%%%%%%%%%%%%%%%%%%%%%%%%%%%%%%%%%%%%%%%%%%%%%%%%%%%%%%%%%%%%%

\clearpage 
%%%%%%%%%%%%%%%%%%%%%%%%%%%%%%%%%%%%%%%%%%%%%%%%%%%%%%%%%%%%%%%%%%%%%%%
%%%%%%%%%%%%%%%%%%%%%%%%%%%%%%%%%%%%%%%%%%%%%%%%%%%%%%%%%%%%%%%%%%%%%%%

\clearpage 
%%%%%%%%%%%%%%%%%%%%%%%%%%%%%%%%%%%%%%%%%%%%%%%%%%%%%%%%%%%%%%%%%%%%%%%
%%%%%%%%%%%%%%%%%%%%%%%%%%%%%%%%%%%%%%%%%%%%%%%%%%%%%%%%%%%%%%%%%%%%%%%

\clearpage 
%%%%%%%%%%%%%%%%%%%%%%%%%%%%%%%%%%%%%%%%%%%%%%%%%%%%%%%%%%%%%%%%%%%%%%%
%%%%%%%%%%%%%%%%%%%%%%%%%%%%%%%%%%%%%%%%%%%%%%%%%%%%%%%%%%%%%%%%%%%%%%%

\clearpage 
%%%%%%%%%%%%%%%%%%%%%%%%%%%%%%%%%%%%%%%%%%%%%%%%%%%%%%%%%%%%%%%%%%%%%%%
%%%%%%%%%%%%%%%%%%%%%%%%%%%%%%%%%%%%%%%%%%%%%%%%%%%%%%%%%%%%%%%%%%%%%%%

\clearpage 
%%%%%%%%%%%%%%%%%%%%%%%%%%%%%%%%%%%%%%%%%%%%%%%%%%%%%%%%%%%%%%%%%%%%%%%
%%%%%%%%%%%%%%%%%%%%%%%%%%%%%%%%%%%%%%%%%%%%%%%%%%%%%%%%%%%%%%%%%%%%%%%

\clearpage 
%%%%%%%%%%%%%%%%%%%%%%%%%%%%%%%%%%%%%%%%%%%%%%%%%%%%%%%%%%%%%%%%%%%%%%%
%%%%%%%%%%%%%%%%%%%%%%%%%%%%%%%%%%%%%%%%%%%%%%%%%%%%%%%%%%%%%%%%%%%%%%%

\end{document}